\documentclass[10pt,twocolumn,twoside]{IEEEtran}
\usepackage{setspace}
\usepackage{stfloats}
\usepackage{float}
\usepackage{cite}
\usepackage{psfig}
\usepackage[dvips]{graphicx}
\usepackage{amsmath}
\usepackage{amssymb}
\usepackage{multirow}
\newcommand{\bb}[1]{{\mathbf{#1}}}
\newcommand{\nn}{\nonumber}

\newcommand{\AW}{\overline}

\newcommand{\AF}{\widehat}

\newcommand{\mysection}[1]{\section{#1}}
\newcommand{\mysubsection}[1]{\subsection{#1}}
\newcommand{\vsp}[1]{}
\newcommand{\np}{}
\begin{document}

\title{STFT with Adaptive  Window Width Based on the Chirp Rate}

\author{Soo-Chang Pei,~\IEEEmembership{Fellow,~IEEE,}
and Shih-Gu Huang
\thanks{Copyright (c) 2012 IEEE. Personal use of this material is permitted. However, permission to use
this material for any other purposes must be obtained from the IEEE by sending a request to
pubs-permissions@ieee.org. This work was supported by the National Science Council, Taiwan, under Contract 98-2221-E-002-077-MY3.

S.~C. Pei is with the Department of Electrical Engineering \& Graduate Institute of Communication Engineering, National Taiwan University, Taipei 10617, Taiwan (e-mail: pei@cc.ee.ntu.edu.tw).

S.-G. Huang is with the Graduate Institute of Communication Engineering, National Taiwan University, Taipei 10617, Taiwan (e-mail: d98942023@ntu.edu.tw).
}
}

\maketitle
\vsp{-1cm}
\begin{abstract}
An adaptive time-frequency representation (TFR) with higher energy concentration usually requires 
higher complexity.
Recently, a
low-complexity adaptive short-time Fourier transform (ASTFT) based on the chirp rate has been proposed.
To enhance the performance, this 
method is substantially modified in this paper: i) because the wavelet transform used for instantaneous frequency (IF) estimation is not signal-dependent, 
a low-complexity ASTFT based on a novel 
concentration measure is addressed; ii)
in order to increase robustness to IF estimation error, the principal component analysis (PCA) replaces the difference operator for calculating the chirp rate; 
and iii) a more robust Gaussian kernel with time-frequency-varying window width is proposed.
Simulation results show that our method has higher energy concentration than the other ASTFTs, especially for multicomponent signals and nonlinear FM signals.
Also, for IF estimation, our method is superior to many other adaptive TFRs 
in low signal-to-noise ratio (SNR) environments.
\end{abstract}

\vsp{-0.25cm}

\mysection{Introduction}\label{sec:Intro}
\vsp{-0.1cm}
\IEEEPARstart{T}ime-frequency (TF) analysis has flourished in various researches and applications in recent years because most signals encountered in practice are not stationary.
TF analysis can reveal comprehensive information about non-stationary signals due to the capability of analyzing a signal in the temporal and spectral domains simultaneously.
Some popular conventional 
TF representations (TFRs) include short-time Fourier transform (STFT), Wigner-Ville distribution (WVD), wavelet transform (WT) \cite{Mallat}, and S-transform \cite{Stockwell}.
In an ideal case, 
a TFR should reveal only the spectral information about the signal occurring at any given time instant.
Accordingly, the main objective of 
a TFR is to provide a more concentrated TF energy distribution without cross terms such that it can resemble as closely as possible to the ideal TFR. 
A variety of more sophisticated and involved 
TFRs have been proposed, such as generalized S-transform \cite{McFadden,Pinnegar}, Hartley S-transform \cite{Pinnegar2},
and Cohen's class 
TFRs using reduced interference distributions (RIDs) \cite{Jeong} or L-class distributions \cite{Stankovic}.
Interested readers can refer to \cite{Sejdic} for an overview of these TFRs.

Researchers believe that no single 
TFR can 
be claimed to have the highest energy concentration for all kinds of signals.
The aforementioned 
TFRs are only appropriate to a limited class of signals or 
require some prior knowledge regarding the signal under analysis.
This explains why there is a growing interest in ``signal-dependent''
TFRs, taking the advantage of the recent development of more powerful computational hardware.
A variety of early developments associated with adaptive
TFRs 
have been 
summarized and cataloged in \cite{Sejdic}.
Numerous recent researches are also proposed in the literature such as adaptive STFT
(ASTFT) \cite{Jaillet,Rudoy,Qaisar,Jiang,Zhong}, adaptive S-transform \cite{Sejdic2,Djurovic,Pei,Lin}, 
adaptive WVD \cite{Jiantao,Ghoraani,Rajshekhar,Tan}, and adaptive
smoothed pseudo WVD (SPWVD) \cite{Behzad}.
To design an adaptive
TFR such that high energy concentration can be achieved,
some methodologies have also been introduced in \cite{Sejdic}, including concentration measures (CMs), reassignment methods and signal optimized kernels/windows.
The last 
one is not taken into account in this paper because 
TFRs based on this approach 
are suitable for a class of signals
rather than all kinds of signals.
A CM has the ability of quantitatively evaluating the TF energy concentration.
In order to achieve the highest energy concentration
(in the sense of the CM), the optimal values of the parameters in a TFR can be obtained by maximizing the 
CM.
However, the main disadvantage of the CM approach is the very high computational complexity.
The CM approach has been used in various TFRs such as the STFT \cite{Jones}, the S-transform \cite{Djurovic,Pei}, the S-method (SM) \cite{Stankovic2}, and the SPWVD \cite{Behzad}.
For each TF point, the reassignment methods calculate the center of gravity of the signal energy around this TF point.
The reassigned TFR is obtained by moving the value of the TFR at each TF point to its corresponding center point.
TFRs based on the reassignment methods have very high energy concentration, but they are computationally expensive and sensitive to noise.
Numerous reassigned TFRs have been proposed such as the reassigned SM \cite{Djurovi2}, the reassigned SPWVD \cite{Auger}, and the reassigned  Wigner-Ville spectrum \cite{Xiao}.

Recently, Zhong and Huang \cite{Zhong} introduced a low-complexity ASTFT based on the 
chirp rate of the signal, i.e. the
first derivative of the instantaneous frequency (IF).
The concept 
is tuning the window width at each time instant such that the signal 
inside the window is quasi-stationary.
Accordingly, a relationship between the window width and the chirp rate 
was addressed:
a wide window is employed as
the IF varies smoothly (chirp rate is small); and
a narrow window is employed as
the IF varies sharply (chirp rate is large).
This 
chirp-rate-based method has the benefit of much lower computational complexity than the CM-based 
methods.
However, the 
TFR utilized for IF estimation in this method is the WT, which is not signal-dependent,
and the difference  operator for calculating the chirp rate is sensitive to IF estimation error.
Besides, the relationship between the window width and the chirp rate is 
not accurate enough, and there is no instruction about how to determine the optimal value of the threshold used in 
this relationship.

To overcome the problems mentioned above, 
this chirp-rate-based method is substantially modified.
First,
a low-complexity CM-based ASTFT is used for IF estimation because it is more suitable for all kinds of signals and more flexible in adjustment between complexity and energy concentration.
Second, the 
principal component analysis (PCA) is introduced for chirp rate estimation since it is less sensitive to the IF estimation error.
Third, Cohen has derived an approximate relationship between the optimal time-varying window width and the chirp rate \cite{Cohen}.
This relationship is more concise and more accurate than that introduced in \cite{Zhong}.
Based on this relationship, 
a Gaussian kernel with time-frequency-varying window width 
is designed by 2D interpolation. 
Accordingly, a new chirp-rate-based ASTFT using this Gaussian kernel is proposed, which is more suitable for nonlinear FM signals and multicomponent signals.

The FFT-based
implementations
of 
the proposed ASTFT are also introduced.
Simulation results show that our method outperforms the CM-based 
ASTFT \cite{Pei} and the chirp-rate-based ASTFT \cite{Zhong} in
both noiseless and noisy environments.
For IF estimation based on TFRs, it is shown that our method is superior to many other adaptive TFRs at low signal-to-noise ratio (SNR) but inferior to the adaptive bilinear TFRs at high SNR.
However, in some applications such as signal analysis and synthesis, our method may be more useful in both low SNR and high SNR environments because it is a linear transform.

This paper is organized as follows.
Section~\ref{sec:ATFA} provides a review of some CM-based adaptive 
S-transforms and the chirp-rate-based ASTFT.
Details of the proposed ASTFT are discussed in Section~\ref{sec:ASTFT}.
Section~\ref{sec:SIM} 
shows the simulation results and comparisons between the proposed 
method and other adaptive TFRs.
The FFT-based
implementations of the proposed 
method are also given in this section.
Finally, conclusions are 
made in Section~\ref{sec:Con}.

\mysection{Adaptive Short-Time Fourier Transforms and Adaptive S-Transforms}\label{sec:ATFA}
This paper focuses on linear TFRs including the STFTs and the S-transforms.
In this section, 
a brief introduction to some CM-based and chirp-rate-based TFRs
is given, the concepts of which will be used in our method.

\vsp{-0.1cm}
\mysubsection{Adaptive STFT and Adaptive S-Transform Based on Concentration Measures}\label{subsec:CM}
The standard S-transform of a signal $x(t)$ is given by
\begin{equation}\label{eq:CM02}
S(t,f) = \int_{ - \infty }^\infty  {x\left( \tau  \right)\frac{{|f|}}{{\sqrt {2\pi } }}{e^{ - \frac{{{f^2}}}{2}{{(t - \tau )}^2}}}{e^{ - j2\pi f\tau }}d\tau },
\end{equation}
where the window kernel is a Gaussian function with standard deviation $\sigma (f) =1/|f|$.
Because the S-transform is not suitable for all kinds of signals, Djurovi\'{c} \emph{et al.} \cite{Djurovic} introduced another variable $p$ to the standard deviation function; that is, $\sigma (f) = 1/|f|^p$.
The modified S-transform is then defined as
\begin{equation}\label{eq:CM04}
{S_p}(t,f) = \int_{ - \infty }^\infty  {x\left( \tau  \right)\frac{{|f{|^p}}}{{\sqrt {2\pi } }}{e^{ - \frac{{{f^{2p}}}}{2}{{(t - \tau )}^2}}}{e^{ - j2\pi f\tau }}d\tau }.
\end{equation}
The optimal value of $p$ at frequency $f$ is obtained by
maximizing concentration measure CM1, which is defined as
\vsp{-0.2cm}
\begin{equation}\label{eq:CM06}
CM1(f,p) = \frac{1}{{\int_{ - \infty }^\infty  {{{\left| \AW{{S_p}(t,f)} \right|}^\alpha }} dt}},
\end{equation}
where $\alpha \in (0,0.25]$.
$\AW{{S_p}(t,f)}$ is the normalized S-transform given by
\begin{equation}\label{eq:CM07}
\AW{{S_p}(t,f)}= \frac{{{S_p}(t,f)}}{{ \int_{ - \infty }^\infty  {{{\left| {{\rm{ }}{S_p}(t,f)} \right|}} dt} }}.
\end{equation}
It is apparent that $1/|f{|^{0.25}} \le {\sigma _{opt}}(f) < 1$ when $f>1$.
Accordingly, this 
TFR somewhat inherits the characteristic of the TF 
localization of the standard S-transform, especially at low frequencies.

A more flexible 
TFR 
should have higher ability to adapt to all kinds of signals, and therefore it seems
\np
unnecessary to set any constraint on the standard deviation of the S-transform.
Accordingly, a more flexible modified S-transform was proposed by Pei and Wang \cite{Pei},
\begin{equation}\label{eq:CM08}
{S_\sigma }(t,f) = \int_{ - \infty }^\infty  {x\left( \tau  \right)\frac{1}{{\sqrt {2\pi } \sigma (f)}}{e^{ - \frac{{{{(t - \tau )}^2}}}{{2{{\left[ {\sigma (f)} \right]}^{\rm{2}}}}}}}{e^{ - j2\pi f\tau }}d\tau },
\end{equation}
where $\sigma(f)$ can be arbitrary positive.
The optimal value of $\sigma(f)$ at frequency $f$ is obtained by maximizing
another concentration measure CM2 defined as
\begin{equation}\label{eq:CM10}
CM2\left(f,\sigma(f)\right) = \int_{ - \infty }^\infty  {{{\left| \AW{{S_\sigma }(t,f)} \right|}^\beta }} dt,
\end{equation}
where $\beta$ is a little 
larger than $1$
and $\AW{{S_\sigma }(t,f)}$ is the normalized S-transform\footnote{Originally, the S-transforms in the CM1 \cite{Djurovic} and the CM2 \cite{Pei} are not normalized.
However, in the Matlab code used in \cite{Pei}, the normalization is employed.
Thus, same normalization method (as shown in (\ref{eq:CM07})) is used in the CM1 and the CM2 here.
In Section~\ref{subsec:VAR}, it will be 
proven that the choice of $\alpha$ in CM1 (or $\beta$ in CM2) would not affect the optimal standard deviation when normalization is employed.}.
The 
modified S-transform in (\ref{eq:CM08}) can be classified as a kind of 
ASTFT because its TF 
localization is no longer relative to that of the S-transform.
More specifically, it can be viewed as an ASTFT with frequency-varying window width.

The main disadvantage of these 
CM-based TFRs is the high computational 
complexity in the optimization process.
Another 
drawback is that for a multicomponent signal, the 
optimal standard deviation obtained from CM1 or CM2 may not be simultaneously optimal for all the components.
This is because 
these CMs
concern the ``total'' energy concentration along the time axis at a certain frequency.

\mysubsection{Adaptive STFT Based on the Chirp Rate}\label{subsec:IFG}
When the window width of the Gaussian kernel is 
time-varying but not frequency-varying, the ASTFT is given by
\vsp{-0.3cm}
\begin{equation}\label{eq:IFG02}
ASTFT_t(t,f) = \int_{ - \infty }^\infty  {x\left( \tau  \right)\frac{1}{{\sqrt {2\pi } \sigma (t)}}{e^{ - \frac{{{{(t - \tau )}^2}}}{{2{{\left[ {\sigma (t)} \right]}^{\rm{2}}}}}}}{e^{ - j2\pi f\tau }}d\tau }.
\end{equation}
Zhong and Huang \cite{Zhong} introduced an algorithm to determine
$\sigma(t)$ for each time instant such that the signal inside the Gaussian window is always quasi-stationary.
This implies that a wide window should be 
employed as the IF
of the signal varies smoothly, while a narrow window should be 
employed as the IF varies sharply.
Accordingly, the window width 
should depend on the chirp rate of the signal, i.e. the first derivative of the IF.
Based on this concept, firstly the IF, $f_{inst}(t)$, is estimated by detecting the ridge of the 
WT of the signal \cite{Delprat,Carmona}.
Then, the chirp rate is obtained from
\begin{equation*}\label{eq:IFG03}
{{f'}_{inst}}(t) = \frac{d}{{dt}}{f_{inst}}(t).
\end{equation*}
The quasi-stationary window width $L_t$ 
is determined by the chirp rate via the following relationship:
\vsp{-0.1cm}\begin{equation}\label{eq:IFG04}
{L_t}=\max_{l}\;2l \quad \textrm{s.t.}\quad \int_{t - l}^{t + l} {\left| {{{f'}_{inst}}(\tau)} \right|}\; d\tau \le \xi.
\end{equation}
$L_t$ is tuned by the threshold $\xi$ such that the integral signal in $ASTFT_t(t,f)$ is  quasi-stationary for every time instant $t$.
If $L_t$ is defined as the full width at half maximum (FWHM) of the Gaussian window, i.e.
\vsp{-0.2cm}\begin{equation}\label{eq:IFG05}
L_t=2\sqrt{2\ln2}\;\sigma(t),
\vsp{-0.3cm}\end{equation}
the standard deviation $\sigma(t)$
is determined from (\ref{eq:IFG05}).
For a discrete signal with sampling interval $\Delta t$, 
the discrete chirp rate at the $k$-th time sampling point is given by
\vsp{-0.1cm}\begin{equation}\label{eq:IFG06}
{{f'}_{inst}}[k] = \frac{{{f_{inst}}[k + 1] - {f_{inst}}[k]}}{{\Delta t}},
\end{equation}
where $f_{inst}[k]$ is the discrete IF.
The relationship in (\ref{eq:IFG04}) toward the discrete signal can be rewritten as
\vsp{-0.1cm}\begin{equation}\label{eq:IFG08}
{L_k} = {\max _l}\;{\kern 1pt} 2l\Delta t\quad \,{\rm{s}}.{\rm{t}}.\quad \,\sum\limits_{m = k - l}^{k + l} \left|{{{f'}_{inst}}[m]}\right|\Delta t \le \xi,
\end{equation}
where $L_k$ is the quasi-stationary window width at the $k$-th sampling point.

The main disadvantages and problems of this 
method are exposed.
First, the WT used for IF estimation in this 
method is not suitable for all kinds of signals.
Second, the difference operator in (\ref{eq:IFG06}) is sensitive to IF estimation error.
Third, the accuracy of the quasi-stationary window width $L_k$ depends on the sampling interval $\Delta t$ and the threshold $\xi$.
The derivation of the optimal value of $\xi$ is not provided in \cite{Zhong}, and this optimal value may be dependent 
on the signal, which leads to higher computational complexity.
Furthermore, the relationships in (\ref{eq:IFG04}) and (\ref{eq:IFG08}) cannot provide the optimal window width, which will be illustrated in Section~\ref{sec:SIM}.
Fourth, for a multicomponent signal,
the quasi-stationary window width is obtained 
from the ``average'' of the
different chirp rates of all the components, and thus not simultaneously the optimal 
for all of them.

\mysection{Adaptive STFT Based on the Concentration Measures and the Chirp Rate}\label{sec:ASTFT}
In our 
method, a generalized ASTFT is introduced to allow 
further control over the 
window width,
\vsp{-0.1cm}\begin{eqnarray}\label{eq:ASTFT02}
&&ASTFT_{tf}(t,f) \nn\\
&&\qquad= \int\limits_{ - \infty }^\infty  {x\left( \tau  \right)\frac{1}{{\sqrt {2\pi } \sigma (t,f)}}{e^{ - \frac{{{{(t - \tau )}^2}}}{{2{{\left[ {\sigma (t,f)} \right]}^{\rm{2}}}}}}}{e^{ - j2\pi f\tau }}d\tau }.
\quad
\end{eqnarray}
The time-frequency-varying standard deviation ${\sigma (t,f)}$ is more suitable for monocomponent nonlinear FM signals 
and multicomponent signals, avoiding the problems of using ${\sigma (f)}$ in (\ref{eq:CM08}) and ${\sigma (t)}$ in (\ref{eq:IFG02}) as mentioned in Section~\ref{sec:ATFA}.
Cohen has derived an approximate relationship between the optimal time-varying window width and the chirp rate \cite{Cohen}.
Based on this relationship, a time-frequency-varying standard deviation is designed by 2D interpolation which will be described later.
A simple and straightforward 
approach for chirp rate estimation is to evaluate the gradients of the IFs of the signal.
There are numerous methods for IF estimation, such as methods based on TFRs, cubic phase function \cite{Oshea}, product high-order ambiguity function \cite{Barbarossa}, and discrete chirp-Fourier transform \cite{Xia}.
In this paper, 
IF is estimated by a novel low-complexity CM-based ASTFT. 
The motivation and details will be described later.

It is apparent that (\ref{eq:ASTFT02}) is equivalent to (\ref{eq:CM08}) when $\sigma(t,f)=\sigma(f)$ and equivalent to (\ref{eq:IFG02}) when $\sigma(t,f)=\sigma(t)$.
In the rest of the paper, our 
method is called
\textbf{ASTFT-tf} for short while the CM-based 
method in (\ref{eq:CM08}) and the chirp-rate-based 
method in (\ref{eq:IFG02}) are called
\textbf{ASTFT-f} and
\textbf{ASTFT-t}, respectively.
For the ease of expressing our 
method, the discrete version of the ASTFT-tf is considered,
\begin{eqnarray}\label{eq:ASTFT04}
&&ASFT_{tf}[m,n]\nn\\
&&\quad= \sum\limits_{l =  - \infty }^\infty  {x[l]\frac{1}{{\sqrt {2\pi } \sigma [m,n]}}{e^{ - \frac{{{{(m - l)}^2}\Delta _t^2}}{{2{{\left( {\sigma [m,n]} \right)}^{\rm{2}}}}}}}{e^{ - j2\pi nl{\Delta _t}{\Delta _f}}}{\Delta _t}},\qquad
\end{eqnarray}
where $\Delta_t$ and $\Delta_f$ are the sampling time interval and sampling frequency interval, respectively.

\begin{figure*}
\begin{center}
\includegraphics[width=18cm]{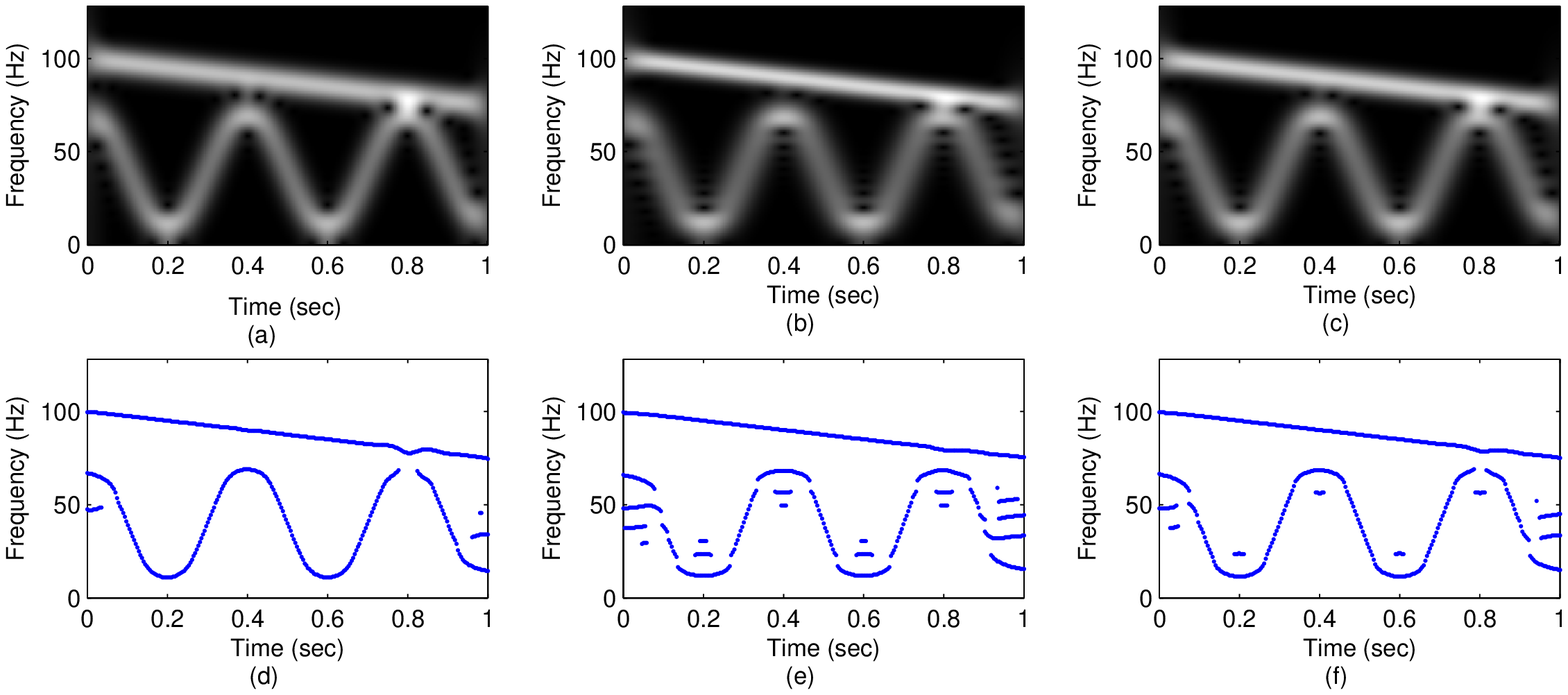} 
\end{center}
\vsp{-0.6cm}
\caption{
The TFRs of $\cos (200\pi t - 20\pi {t^2}) + \cos (4\pi \sin (5\pi t) + 80\pi t)$ and the detected ridges:
(a) CM3-based ASTFT; (b) CM4-based ASTFT; (c) CM5-based ASTFT; (d) detected ridges of the CM3-based ASTFT; (e) detected ridges of the CM4-based ASTFT; and (f) detected ridges of the CM5-based ASTFT.
In this example,
the CM3 performs best; however, it may be the worst for some signals such as the signal in Fig.~\ref{fig:CM2}. Thus,
the CM5 is preferred for IF estimation because it is a compromise between
the CM3 and
the CM4.
}\label{fig:CM1}
\vsp{-0.1cm}
\end{figure*}
\begin{figure*}
\begin{center}
\includegraphics[width=18cm]{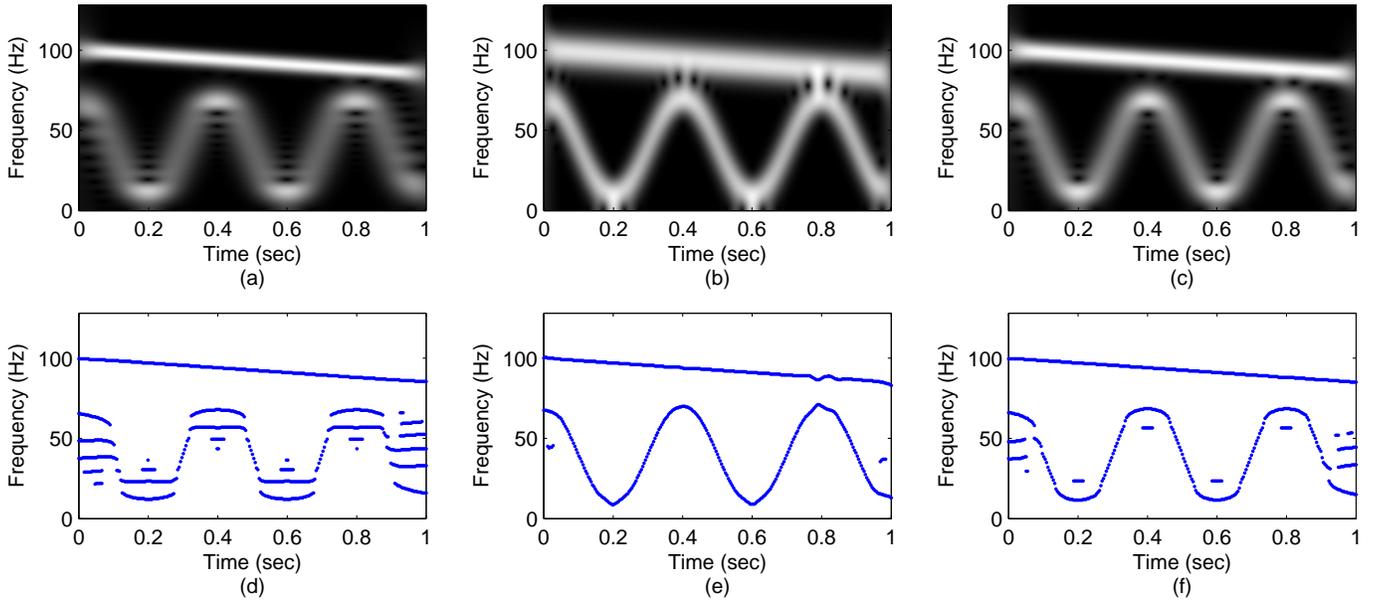} 
\end{center}
\vsp{-0.6cm}
\caption{
The TFRs of $\cos (200\pi t - 10\pi {t^2}) + \cos (4\pi \sin (5\pi t) + 80\pi t)$ and the detected ridges:
(a) CM3-based ASTFT; (b) CM4-based ASTFT; (c) CM5-based ASTFT; (d) detected ridges of the CM3-based ASTFT; (e) detected ridges of the CM4-based ASTFT; and (f) detected ridges of the CM5-based ASTFT.
In this example,
the CM4 performs best; however, it may be the worst for some signals such as the signal in Fig.~\ref{fig:CM1}. Thus,
the CM5 is preferred
for IF estimation because it is a compromise between
the CM3 and
the CM4.
}\label{fig:CM2}
\vsp{-0.7cm}
\end{figure*}

\mysubsection{Instantaneous Frequency Estimation Using a Low-Complexity CM-Based ASTFT}\label{subsec:IFE}
An overview of IF estimation methods based on
TFRs is 
presented in \cite{Boashash,Boashash2,Sejdic}.
It has been known that 
TFRs can concentrate the energy of the signal at and around the ridges in the TF plane.
Therefore, the ridges of the TFR having higher energy concentration would approximate more closely to the exact IFs of the signal \cite{Boashash,Boashash2,Cohen,Chandre,Rankine}.
To design an adaptive TFR with high energy concentration, possible approaches include the CMs, the reassignment methods and the signal optimized kernels/windows which have been 
mentioned in Introduction.
The last one is not considered here because the 
TFRs based on this approach 
are suitable for a class of signals rather than all kinds of signals.
In this paper, the CM approach is adopted because it is less sensitive to noise than the reassignment methods (see Section~\ref{subsec:Noise}).
Besides, the CM approach is more flexible, allowing an adjustment between complexity and energy concentration.

Since 
the purpose of this paper is to design a chirp-rate-based
``ASTFT'', the CM-based
``ASTFT'' rather than other CM-based TFRs is preferred for IF estimation in order to reduce hardware cost.
Generally speaking, a more involved 
TFR with higher energy concentration usually follows a more accurate IF 
estimate.
Fortunately, the main objective of our method is not to obtain the exact IFs.
Small IF estimation error is tolerable, and partial serious estimation error would only induce partial performance loss (see the simulation in Section~\ref{subsec:Error}).
Therefore, low complexity is the top priority, followed by energy concentration.
Recall the CM1 in (\ref{eq:CM06}) and the CM2 in (\ref{eq:CM10}).
To reduce the complexity, 
the following modifications are made:
\begin{itemize}
\item Instead of finding the optimal $\sigma(f)$ for each frequency, we find the optimal $\sigma$ for the entire TF plane.
    Accordingly, 
    CM optimization is performed only once for the parameter $\sigma$.
    In the following, ${X_\sigma }[m,n]$ denotes the ASTFT using $\sigma$ for the entire TF plane.

\item The modified CMs intended to measure the energy concentration for all ${X_\sigma }[m,n]$ observations.
    Nevertheless, for reducing complexity, only part of the observations is concerned. When ${X_\sigma }[p\tilde m,p\tilde n]$'s are concerned, only $\dfrac{1}{p^2}$ of the observations are required to be computed in the optimization process.

\item The optimal value of $\sigma$ is chosen from a 
limited set $\left\{ {{\sigma _1},{\sigma _2}, \ldots ,{\sigma _L}} \right\}$, where $L$ is not large.
\end{itemize}
The flexibility is realized by adjusting the values of $p$ and $L$.
Larger $p$ and smaller $L$ 
can reduce the complexity at the cost of performance loss.

Based on the above modifications, the discrete versions of
the modified CM1 and
the modified CM2 (denoted by CM3 and CM4) are 
\begin{eqnarray}\label{eq:IFE04}
CM3[\sigma ] = \frac{1}{{\sum\limits_{\tilde m} {\sum\limits_{\tilde n} {{{\left| \AW{{X_\sigma }[p\tilde m,p\tilde n]} \right|}^\alpha }} } }},\quad 0 < \alpha  < 1,\\
CM4[\sigma ] = \sum\limits_{\tilde m} {\sum\limits_{\tilde n} {{{\left| \AW{{X_\sigma }[p\tilde m,p\tilde n]} \right|}^\beta }} } ,\quad\beta  > 1,
\end{eqnarray}
where $\AW{{X_\sigma }[p\tilde m,p\tilde n]}$ is the normalized STFT which is similar to the discrete version of (\ref{eq:CM07}).
%
%
Effects of the CM3 and the CM4 on energy concentration and IF estimation are shown in Figs.~\ref{fig:CM1} and \ref{fig:CM2} 
with $\alpha=0.1$, $\beta=5$, $p=4$ and $L=64$.
The signals utilized in these two examples are the same except that the chirp rates of the linear FM components are different.
It is
shown that both maximizing the CM3 and maximizing the CM4 have the ability to 
enhance the energy concentration.
The difference is that the former induces lower total energy while the latter induces higher total energy.
Observing the 
detected ridges (i.e. estimated IFs) in these examples, the CM3 provides more satisfactory IF estimation for the signal in Fig.~\ref{fig:CM1}, while the CM4 is more suitable for the signal in Fig.~\ref{fig:CM2}.
Therefore, 
a new CM which is a compromise between the CM3 and the CM4
is introduced,
denoted as CM5,
\begin{equation}\label{eq:IFE06}
CM5[\sigma ] = \frac{{{{\left( {\sum\limits_{\tilde m} {\sum\limits_{\tilde n} {{{\left| \AW{{X_\sigma }[p\tilde m,p\tilde n]} \right|}^\beta }} } } \right)}^{1/\beta }}}}{{{{\left( {\sum\limits_{\tilde m} {\sum\limits_{\tilde n} {{{\left| \AW{{X_\sigma }[p\tilde m,p\tilde n]} \right|}^\alpha }} } } \right)}^{1/\alpha }}}},\quad {\mkern 1mu} 0 < \alpha  < 1 < {\mkern 1mu} \beta .
\end{equation}
Effect of the CM5 on energy concentration and IF estimation 
is depicted in Figs.~\ref{fig:CM1} and \ref{fig:CM2}.
It is apparent that the CM5-based 
ASTFT has performance between the CM3-based and the CM4-based 
ASTFTs.

For a monocomponent signal, the IF can be easily estimated via detecting the ridge of the 
TFR, i.e. detecting the positions of the maximal energy along the frequency axis at every time instant.
At the $m$-th sampling time, the IF is given by
${f_{inst}}[m] = {f_{inst}}(m\Delta_t)\approx n_m\Delta_f$
where $n_m$ is determined by ridge detection,
$n_m =\arg \mathop {\max }\limits_n \left| {{X_\sigma }[m,n]} \right|$.
If the signal has multiple components, there would be multiple
local maxima along the frequency axis at every time instant.

\mysubsection{Chirp Rate Estimation Using Principal Component Analysis}\label{subsec:CRE}
For lack of prior knowledge of the exact IFs of the signal, some undesired ridges (see Figs.~\ref{fig:CM1}(f) and \ref{fig:CM2}(f)) would also be detected.
Thanks to some postprocess, the ridges with too short length or too small energy can be easily eliminated.
To evaluate the chirp rate, the difference operator in (\ref{eq:IFG06}) is sensitive to IF estimation error, especially in noisy environments.
Thus, in order to increase robustness to the IF estimation error, 
the 
PCA \cite{Dunteman,Jolliffe} is used to obtain the coarse estimate of the chirp rate.
To obtain ${{f'}_{inst}}[m]$ for some $m$, the estimated IF ${{f}_{inst}}[m]$ and its nearby $2K$ estimated IFs are utilized.
Define time variable $T$ and frequency variable $F$ with $2K+1$ measurements $(T_i,F_i)$'s, i.e
($(m - K){\Delta _t}$, ${f_{inst}}[m - K]$), ($(m - K+1){\Delta _t}$, ${f_{inst}}[m - K+1]$), $\ldots$, ($(m + K){\Delta _t}$, ${f_{inst}}[m + K]$).
Then, the slope of the first principal component vector of this set of measurements 
approximates to ${{f'}_{inst}}[m]$.
The covariance matrix of 
the 2-dimensional measurements is given by
\begin{equation}\label{eq:IFE14}
{\bf{C}} =
\begin{bmatrix}
{{C_{TT}}}\;&{{C_{TF}}}\\
{{C_{FT}}}\;&{{C_{FF}}}
\end{bmatrix} \buildrel \Delta \over =
\begin{bmatrix}
{{\mathop{\rm cov}} (T,T)}\;&{{\mathop{\rm cov}} (T,F)}\\
{{\mathop{\rm cov}} (F,T)}\;&{{\mathop{\rm cov}} (F,F)}
\end{bmatrix}.
\end{equation}
Perform the eigenvalue decomposition to $\bf{C}$.
The first principal component vector is equivalent to the eigenvector $\bb e_1=[e_{11},\; e_{12}]^T$ corresponding to the largest eigenvalue $\lambda_1$.
Because ${{f'}_{inst}}[m]$ is 
approximated 
by the slope of $\bb e_1$, we have
\begin{equation}\label{eq:IFE12}
{{f'}_{inst}}[m]
\approx \frac{{{e_{12}}}}{{{e_{11}}}}=\frac{\lambda_1-C_{TT}}{C_{TF}},
\end{equation}
where
\begin{equation}\label{eq:IFE15}
\lambda_1=\frac{C_{TT}+C_{FF}+\sqrt{(C_{TT}+C_{FF})^2-4C_{TF}C_{FT}}}{2}.
\end{equation}
The difference operator in (\ref{eq:IFG06}) can be viewed as a special case of the PCA, using only two IF measurements (i.e. ${{f}_{inst}}[m]$ and ${{f}_{inst}}[m+1]$) to evaluate the chirp rate ${{f'}_{inst}}[m]$.
For noisy signal with lower SNR, more IF measurements (larger $K$) should be used to increase robustness to IF estimation error.
Since the signal may have multiple components, before calculating the chirp rates, ridge curve tracing is necessary
in order to separate the 
estimated IFs into several sets corresponding to different components.

\mysubsection{Optimal Standard Deviation of the Gaussian Window Versus the Chirp Rate}\label{subsec:VAR}
Cohen has derived an approximate relationship between the optimal window width and the chirp rate for purely frequency modulated signals \cite{Cohen}; that is, if
the signal is of the form $x(t)=\exp(j\varphi(t))$, the
optimal window width $T_t^2$ 
can be approximated by
\begin{equation}\label{eq:VAR0000}
T_t^2 \approx \frac{1}{{2\left| {\varphi ''(t)} \right|}}=\frac{1}{{4\pi \left| {f'_{inst}(t)} \right|}},
\end{equation}
where $f'_{inst}(t)$ is the chirp rate.
If the window is a Gaussian function with time-varying standard deviation $\sigma(t)$, 
the window width is given by $T_t^2=\sigma^2(t)/2$ which yields
\begin{equation}\label{eq:VAR0001}
\sigma^2(t) \approx \frac{1}{{2\pi \left| {f'_{inst}(t)} \right|}}.
\end{equation}
Consider the simplest case that $x(t)$ is a linear FM signal, $x(t) = \exp \left( {j2\pi (a{t^2}/2 + bt)} \right)$. The chirp rate is $f'_{inst}(t)=a$,
and the relationship in (\ref{eq:VAR0001}) can be rewritten as
follows ``without any approximation'',
\begin{equation}\label{eq:VAR0003}
\sigma^2(t)=\sigma^2=\frac{1}{{2\pi \left| a \right|}}.
\end{equation}
The derivation of (\ref{eq:VAR0001}) is complicated, and thus we directly prove (\ref{eq:VAR0003}) by analyzing the TF energy distribution of the linear FM signal.
Since the chirp rate is a constant for the linear FM signal, it is reasonable to use $\sigma(t,f)=\sigma$ in (\ref{eq:ASTFT02}).
The envelope of the corresponding ASTFT-tf is given by
\begin{eqnarray}\label{eq:VAR000}
&&\left| ASTFT_{tf}(t,f) \right|\nn\\
&&\qquad\qquad=\left( {1 + 4{\pi ^2}{\sigma ^4}{a^2}} \right)^{ - \frac{1}{4}}}{e^{ - \frac{{2{\pi ^2}{\sigma ^2}}}{{1 + 4{\pi ^2}{\sigma ^4}{a^2}}}{{\left( {f - b - at} \right)}^2}}.\quad
\end{eqnarray}
A detailed derivation of the above equation is given in APPENDIX~\ref{App:A}.
At any time instant, the envelope shown in (\ref{eq:VAR000}) is a Gaussian function of $f$ with variance $\eta(\sigma ^2)$ given by
\begin{equation}\label{eq:VAR002}
\eta (\sigma ^2)= \frac{{1 + 4{\pi ^2}{a^2}{\left(\sigma ^2\right)^2}}}{4{\pi ^2}(\sigma ^2)}.
\end{equation}
In order to achieve the highest energy concentration, the variance $\eta (\sigma ^2)$ should be minimized.
$\eta (\sigma ^2)$ is strictly convex because its second derivative is positive as $\sigma ^2>0$.
Accordingly, 
a global minimum of $\eta (\sigma ^2)$ 
occurs when $\dfrac{d}{{d{\sigma ^2}}}\;\eta (\sigma ^2)=0$.
The optimal standard deviation is then given by
\begin{equation}\label{eq:VAR010}
\sigma^2_{opt} = \sqrt {\frac{1}{{4{\pi ^2}{a^2}}}}  = \frac{1}{{2\pi }}\frac{1}{{\left| a \right|}}.
\end{equation}
At any frequency, the envelope shown in (\ref{eq:VAR000}) is a Gaussian function of $t$.
Similarly, 
$\sigma^2_{opt}$ in (\ref{eq:VAR010}) also leads to the minimal value of the variance of the Gaussian function.

In the following, 
the relationship
shown in (\ref{eq:VAR0003}) and (\ref{eq:VAR010})
is verified from the point of view of CMs.
Substituting $ASTFT_{tf}(t,f)$ for $S_\sigma(t,f)$ in the CM2 in (\ref{eq:CM10}), 
the CM2 can be rewritten as
\begin{eqnarray*}\label{eq:VAR012}
CM2\left(f,\sigma\right) &=& \int_{ - \infty }^\infty  {{{\left| \AW{ASTFT_{tf}(t,f)} \right|}^\beta }} dt\nn\\
&=&\int_{ - \infty }^\infty  {{{\left| {\frac{{ASTF{T_{tf}}(t,f)}}{{\int_{ - \infty }^\infty  {\left| {ASTF{T_{tf}}(t,f)} \right|} dt}}} \right|}^\beta }} dt.
\end{eqnarray*}
For a linear FM signal, the envelope of the ASTFT-tf has depicted in (\ref{eq:VAR000}), and therefore the above formula can be simplified as
\begin{eqnarray}\label{eq:VAR014}
CM2\left( {f,\sigma } \right) &=& \frac{{{{\left( {1 + 4{\pi ^2}{\sigma ^4}{a^2}} \right)}^{\frac{1}{2} - \frac{\beta }{4}}}{{\left( {2\pi {\sigma ^2}{a^2}\beta } \right)}^{ - \frac{1}{2}}}}}{{{{\left( {1 + 4{\pi ^2}{\sigma ^4}{a^2}} \right)}^{\frac{\beta }{4}}}{{\left( {2\pi {\sigma ^2}{a^2}} \right)}^{ - \frac{\beta }{2}}}}} \nn\\
&=& {\beta ^{ - \frac{1}{2}}}{\left( {\frac{1}{{2\pi {\sigma ^2}{a^2}}} + 2\pi {\sigma ^2}} \right)^{\frac{{1 - \beta }}{2}}}.
\end{eqnarray}
The maximum occurs when
\begin{equation}\label{eq:VAR016}
\frac{d}{{d({\sigma ^2})}}\left( {\frac{1}{{2\pi {\sigma ^2}{a^2}}} + 2\pi {\sigma ^2}} \right) =  - \frac{{2\pi {a^2}}}{{{{\left( {2\pi {\sigma ^2}{a^2}} \right)}^2}}} + 2\pi  = 0
\end{equation}
which leads to the same result as in (\ref{eq:VAR0003}) and (\ref{eq:VAR010}).
Similarly, 
the CM1 in (\ref{eq:CM06}) can also be used to verify the relationship\footnote{If normalization is not used, the CM1 would yield $\sigma _{opt}^2 = \frac{1}{{2\pi \left| a \right|\sqrt {1 - \alpha } }}$ while the CM2 would yield $\sigma _{opt}^2 \to 0$.}.
When the signal under analysis is a
highly nonlinear FM signal (the chirp rate is time-varying), the optimal $\sigma(t)$ would involve 
not only $\varphi''(t)$ but also $\varphi'''(t), \varphi^{(4)}(t),\ldots, \varphi^{(\infty)}(t)$
\cite{Cohen}.
It can be predicted that CMs involve
$\varphi''(t), \varphi'''(t), \varphi^{(4)}(t),\ldots, \varphi^{(\infty)}(t)$ and $\sigma^2(t), \sigma^3(t),\ldots, \sigma^{\infty}(t)$.
Therefore, it is more practical to use the approximate relationship depicted in (\ref{eq:VAR0001}).

\begin{figure*}[t]
\begin{center}
\includegraphics[width=18cm]{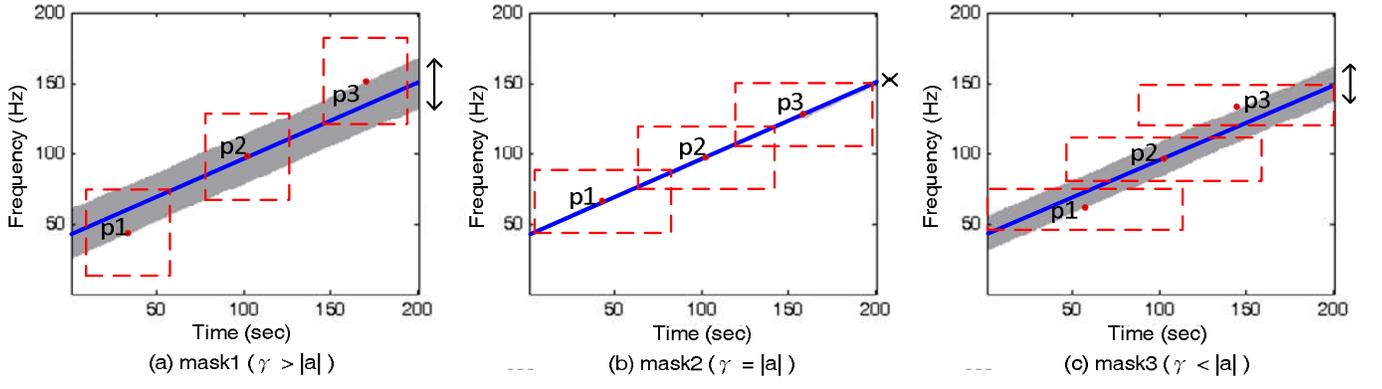}
\end{center}
\vsp{-0.6cm}\caption{
The ideal TFR
(solid straight lines) of a linear FM signal with chirp rate $a$ and three uniform TF masks
(dashed rectangles) with different height-to-width ratios $\gamma$'s: (a) mask1 ($\gamma>|a|$), (b) mask2 ($\gamma=|a|$), and (c) mask3 ($\gamma<|a|$).
The gray block in each sub-figure represents the region 
having the highest envelope of the convolution of the ideal TFR with the 
mask.
This region is equivalent to the ideal TFR as mask2 ($\gamma=|a|$) is used.
} \label{fig:mask1}
\vsp{0cm}\end{figure*}

\begin{figure*}[t]
\begin{center}
\includegraphics[width=18cm]{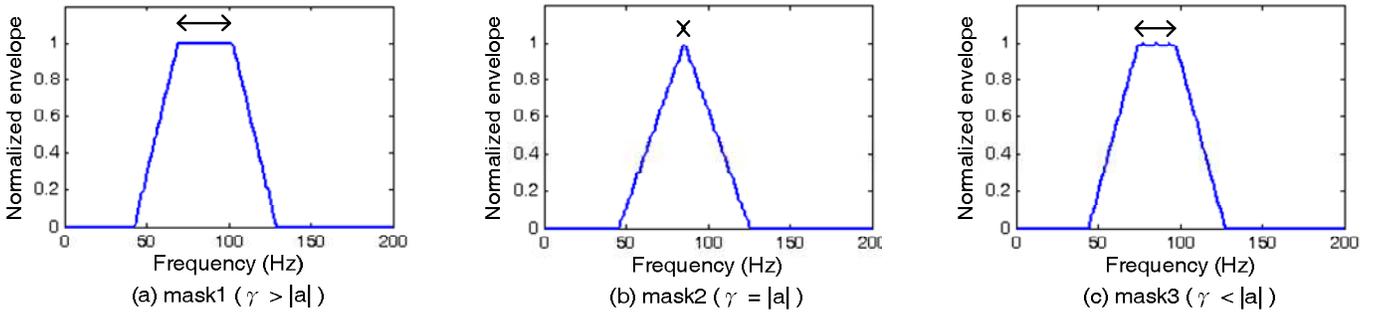}
\end{center}
\vsp{-0.6cm}\caption{
The normalized envelopes of the convolutions of the ideal TFR with the three kinds of TF masks shown in Fig.~\ref{fig:mask1} at $t=80$ sec.: ((a) mask1 ($\gamma>|a|$), (b) mask2 ($\gamma=|a|$), and (c) mask3 ($\gamma<|a|$).
Mask2 ($\gamma=|a|$) can
yield the highest energy concentration.
} \label{fig:mask2}
\vsp{-0.6cm}\end{figure*}

\mysubsection{Time-Frequency-Varying Standard Deviation Using 2D Interpolation}\label{subsec:Interp}
When analyzing a nonlinear FM signal or a multicomponent signal, time-frequency-varying window width is preferred to achieve higher energy concentration than time-varying window width.
Therefore, 
the relationship between the optimal standard deviation and the
chirp rate is examined from another point of view.
A Gaussian kernel $w(t)$ with standard deviation $\sigma$ and its Fourier transform $W(f)$ are given by
\begin{eqnarray}\label{eq:VAR02}
w(t) = \frac{1}{{\sqrt {2\pi } \sigma }}\;{e^{ - \frac{{{t^2}}}{{2{\sigma ^{\rm{2}}}}}}}\;,\quad
W(f) = \sqrt {2\pi } \sigma\; {e^{ - 2{\pi ^2}{\sigma ^{\rm{2}}}{f^2}}}.
\end{eqnarray}
The temporal and spectral spreads of the kernel function are respectively defined as:
\begin{eqnarray}\label{eq:VAR04}
\delta _t^2&=&\frac{{{w_2} - w_1^2}}{{{w_0}}} = \frac{{{\sigma ^2}}}{2},\quad{w_i} = \int_{ - \infty }^\infty  {{t^i}{{\left| {w(t)} \right|}^2}dt} ,\\
\delta _f^2&=&\frac{{{W_2} - W_1^2}}{{{W_0}}} = \frac{1}{{8{\pi ^2}{\sigma ^2}}},\,{W_i} = \int_{ - \infty }^\infty  {{f^i}{{\left| {W(f)} \right|}^2}df}.
\end{eqnarray}
The spreads are sometimes indicated with the Heisenberg box \cite{Mallat}.
In the TF plane, the Gaussian kernel can be deemed as a two-dimensional (2D) mask, i.e. a box with time spread $\delta_t$  and frequency spread $\delta_f$.
If the FWHM
in (\ref{eq:IFG05}) is employed, the Gaussian mask has width $2\sqrt{2\ln2}\;\delta_t$  
and height $2\sqrt{2\ln2}\;\delta_f$, and the height-to-width ratio $\gamma$ is given by
\begin{equation}\label{eq:VAR06}
\gamma=\frac{2\sqrt{2\ln2}\;\delta _f}{2\sqrt{2\ln2}\;\delta _t}=\frac{\delta _f}{\delta _t}=\frac{1}{2\pi\sigma^2}.
\end{equation}
This equation implies that the standard deviation $\sigma$ of the Gaussian kernel can be determined by the height-to-width ratio $\gamma$ of the 2D Gaussian mask.
The TFR can be deemed as the convolution of the ideal TFR with the 2D Gaussian mask.
Therefore, the problem is how to tune the shape (controlled by $\gamma$) of the mask for every TF point such that the TFR has energy as concentrated on its ridges as possible.

To express the notion of the answer, 
the problem is simplified by considering that the 2D mask is ``uniform''.
For a discrete signal
consisting of only one linear FM component, 
\vsp{-0.2cm}
\begin{equation}\label{eq:VAR07}
x[m]=\exp \left( {j2\pi (a(m\Delta_t)^2/2 + bm\Delta_t)} \right),
\end{equation}
the exact chirp rate is a constant, i.e. ${f'}_{inst}[m] = a$.
The ideal TFR of the signal is shown in Fig.~\ref{fig:mask1} (
solid straight lines).
To examine the energy concentration of the convolution of the ideal TFR with the 2D uniform mask,
three kinds of masks with height-to-width ratios $\gamma>|a|$, $\gamma=|a|$ and $\gamma<|a|$  are utilized, as depicted in Figs.~\ref{fig:mask1}(a), \ref{fig:mask1}(b) and \ref{fig:mask1}(c)
(dashed rectangles), respectively.
The gray block in each sub-figure of Fig.~\ref{fig:mask1} represents the region of the TF points having the highest envelope of the convolution.
It is obvious that the region is exactly the distribution of the ideal TFR when $\gamma=|a|$.
Because the signal is a linear FM signal, the distributions of the convolution along the frequency axis at all time instants are similar.
Therefore, Fig.~\ref{fig:mask2} only shows the normalized envelopes of the convolutions 
at $t=80$ (sec.).
The envelope is nonzero between $43$Hz and $128$Hz for all the three masks; however, the envelope is the most concentrated when the mask with $\gamma=|a|$ is utilized.

For a 2D Gaussian mask, which is nonuniform,
the height-to-width
ratio $\gamma=|a|$ is also the optimal choice, but the difference of the concentration levels as shown in Fig.~\ref{fig:mask2} would be not so significant.
According to (\ref{eq:VAR06}), the optimal standard deviation $\sigma_{opt}$ is then determined by
\begin{equation}\label{eq:VAR08}
\sigma^2_{opt} =\frac{1}{{2\pi }}\frac{1}{{\gamma}}=\frac{1}{{2\pi }}\frac{1}{{\left|a\right|}}.
\end{equation}
Since this result is equivalent to that in (\ref{eq:VAR0003}) and (\ref{eq:VAR010}), it is feasible to determine the optimal standard deviation from the shape of the 2D mask.
Note that the chirp rate 
may be $0$ or $\pm\infty$
in some cases,
and thus upper bound $\sigma_{max}$ and lower bound $\sigma_{min}$ of the standard deviation should be defined.
For 
instance,
$2\sqrt{2\ln2}\;\sigma_{max}$ can be set equal to the signal length.

\begin{figure}[t]
\begin{center}
\includegraphics[width=8.3cm]{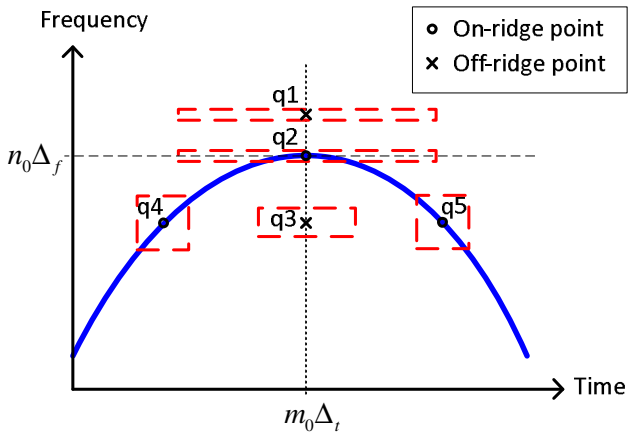}
\end{center}
\vspace{-0.4cm}\vsp{-0.5cm}\caption{
The mask dilation strategy for a nonlinear FM signal.
The solid line is the ideal TFR.
For the
on-ridge points (q2, q4 and q5), the height-to-width ratios $\gamma$'s of the masks are equal to the absolute values of the chirp rates.
For the off-ridge points (q1 and q3), the mask at q1 
should be the same as that at q2; however, $\gamma$ of the mask at q3 should be in-between those at q2, q4 and q5
to avoid overlapping with the ideal TFR.
} \label{fig:mask3}
\vspace{-0.4cm}\vsp{-0.6cm}\end{figure}

\begin{figure*}[b]
\begin{center}
\includegraphics[width=17.5cm]{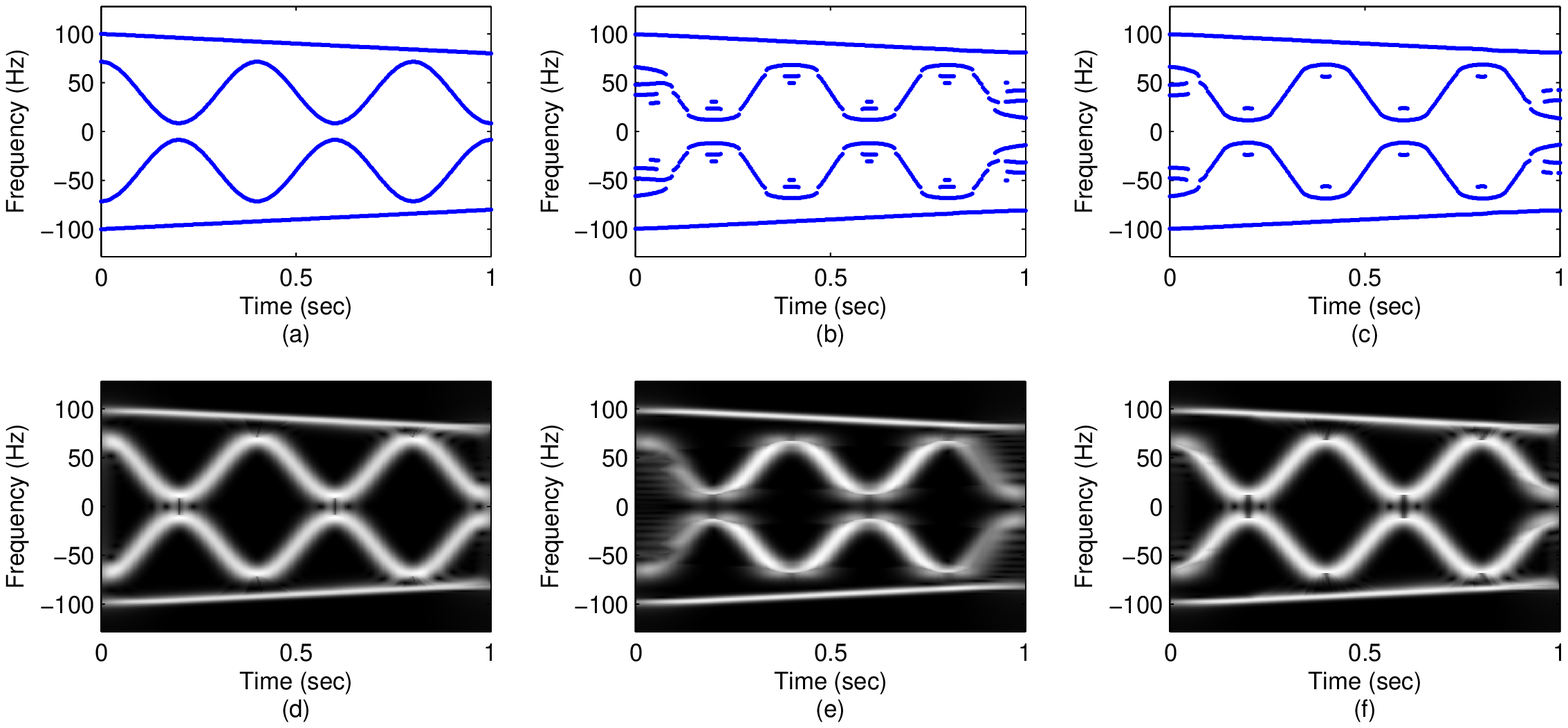}
\end{center}
\vspace{-0.3cm}\vsp{-0.5cm}\caption{
Effect of IF estimation error on the performance of the ASTFT-tf: (a) exact ridges; (b) detected ridges from the CM3-based ASTFT (c) detected ridges from the CM5-based ASTFT; (d)
ASTFT-tf using the ridges in (a); (e)
ASTFT-tf using the ridges in (b); and (f)
ASTFT-tf using the ridges in (c).
Subfigure (e) shows that partial serious IF estimation error would only induce partial ASTFT performance loss. Subfigure (f) shows that small IF estimation error is tolerable.
} \label{fig:SIM1}
\vsp{-0.6cm}\end{figure*}

Consider the more complicated case that the signal under analysis
consists of multiple components or a nonlinear FM component.
Because the chirp rate is no longer a constant, 
${f'}_{inst}[m,n]$ is defined as:
\begin{itemize}
\item If $(m\Delta_t,n\Delta_f)$ is on the ridge (called an on-ridge point), ${f'}_{inst}[m,n]$ is define as the chirp rate of the component occurring at this TF point.
\item If $(m\Delta_t,n\Delta_f)$ is off the ridge (called an off-ridge point), ${f'}_{inst}[m,n]$ is undefined.
\end{itemize}
The ideal TFR of a monocomponent nonlinear FM signal is depicted in Fig.~\ref{fig:mask3}.
The points q2, q4 and q5 are on-ridge points, while q1 and q3 are off-ridge points.
According to Cohen's derivation in (\ref{eq:VAR0001}), the 
optimal standard deviation of the on-ridge point with chirp rate ${f'}_{inst}[m,n]$ can be 
approximated by
\begin{equation}\label{eq:VAR10}
{\sigma ^2_{opt}}[m,n] \approx  \frac{1}{{2\pi }} \cdot {\mkern 1mu} \frac{1}{\left|{f'}_{inst}[m,n]\right|}.
\end{equation}
The problem is how to determine the optimal standard deviations for the off-ridge points.
Observe the shapes of the Gaussian masks of the on-ridge points q2, q4 and q5, as shown in Fig.~\ref{fig:mask3}.
To achieve high energy concentration, the height-to-width ratio of the mask at q1 
should be the same as 
that at q2; however, the height-to-width ratio of the mask at q3 should be in-between 
those at q2, q4 and q5
to avoid overlapping with the ideal TFR.
This implies that at time instant $m_0\Delta_t$ in Fig. 5, applying a single value $\sigma[m_0]$ to the entire frequency band is worse than using $\sigma[m_0,n]$.
Similarly, at frequency $n_0\Delta_f$, applying a single value $\sigma[n_0]$ to the entire time interval would be worse than using $\sigma[m,n_0]$.
For the purpose of low complexity,
2D interpolation
is employed to obtain the $\gamma$'s (i.e. ${f'}_{inst}[m,n]$'s) for all the off-ridge points.
Once all the ${f'}_{inst}[m,n]$'s are determined, the approximate optimal standard deviations for all the TF points can be obtained from (\ref{eq:VAR10}).

In our simulations, 2D triangle-based linear interpolation on the ${f'}_{inst}[m,n]$ is utilized.
Although this interpolation
method may not be the optimal, 
it can 
achieve higher energy concentration among some well known interpolations on the  ${f'}_{inst}[m,n]$, the $\tan^{-1}({f'}_{inst}[m,n])$ or the $1/(2\pi)/{f'}_{inst}[m,n]$:
nearest neighbor interpolation, triangle-based linear interpolation, triangle-based cubic interpolation and MATLAB 4 griddata method.
There is always a tradeoff between energy concentration and complexity.
Therefore, it is impractical to design an ASTFT with enormous amount of computation even though it has the highest energy concentration.
Although the proposed technique is not the best for energy concentration, it has a great advantage in terms of low complexity.

\mysection{Simulation Results}\label{sec:SIM}
In this section, several experiments are given to compare the performance of the ASTFT-f, the ASTFT-t and the ASTFT-tf, which in turns represent the CM-based ASTFT \cite{Pei} introduced in Section~\ref{subsec:CM}, the chirp-rate-based ASTFT \cite{Zhong} introduced in Section~\ref{subsec:IFG}, and our 
method proposed in Section~\ref{sec:ASTFT}.
We also examine the performance of IF 
estimators based on the
ASTFT-tf and other adaptive TFRs, including some popular adaptive bilinear TFRs, in noisy environments.
In these experiments, $\alpha=0.1$, $\beta=5$, $p=4$ and $L=64$ are utilized in the CM5 
of the
ASTFT-tf.
As mentioned 
before, in the
original ASTFT-t, the WT used for IF estimation is not signal-dependent, and the difference operator used for calculating the chirp rate is sensitive to IF estimation error.
Therefore, 
the chirp rate obtained from the ASTFT-tf is applied to the ASTFT-t in all the following simulations.

\mysubsection{Effect of IF Estimation Error on the Performance of the ASTFT-tf}\label{subsec:Error}
The standard deviation in the ASTFT-tf is dependent on the chirp rate of the signal.
Therefore, accuracy of IF estimation would influence the performance.
In this paper, a low-complexity CM5-based ASTFT is adopted for IF estimation in the ASTFT-tf.
To analyze the 
effect of the IF estimation error on the 
energy concentration,
the ASTFT-tf with perfect IF estimation is compared with the original ASTFT-tf (using the CM5) and the ASTFT-tf substituting the CM5 for the CM3.
Consider a synthetic signal given by
\vsp{-0.1cm}\begin{equation*}\label{eq:SIM02}
x(t) = \cos \left( {200\pi t - 20\pi {t^2}} \right) + \cos \left( {4\pi \sin (5\pi t) + 80\pi t} \right),
\vsp{-0.1cm}\end{equation*}
with $\Delta_t=1/256$ and $\Delta_f=1$.
The ridges shown in Fig.~\ref{fig:SIM1}(a) are the exact IFs, and the corresponding ASTFT-tf is depicted in Fig.~\ref{fig:SIM1}(d).
The ridges shown in Figs.~\ref{fig:SIM1}(b) and \ref{fig:SIM1}(c) are respectively 
obtained from the IF 
estimation methods based on the CM3 and
the CM5.
The ASTFT-tf corresponding to the CM3 and the ASTFT-tf corresponding to the CM5 are depicted in Figs.~\ref{fig:SIM1}(e) and \ref{fig:SIM1}(f), respectively.
It is shown that higher IF estimation error would lead to lower energy concentration.
By comparing Figs.~\ref{fig:SIM1}(d) and \ref{fig:SIM1}(f), 
the performance loss induced by the CM5-based IF estimation is tolerable.
This explains why the low-complexity CM5-based ASTFT rather than other more 
involved methods is adopted for IF estimation.

\begin{figure}
\vspace{0.5cm}\begin{center}
\includegraphics[width=6.75cm]{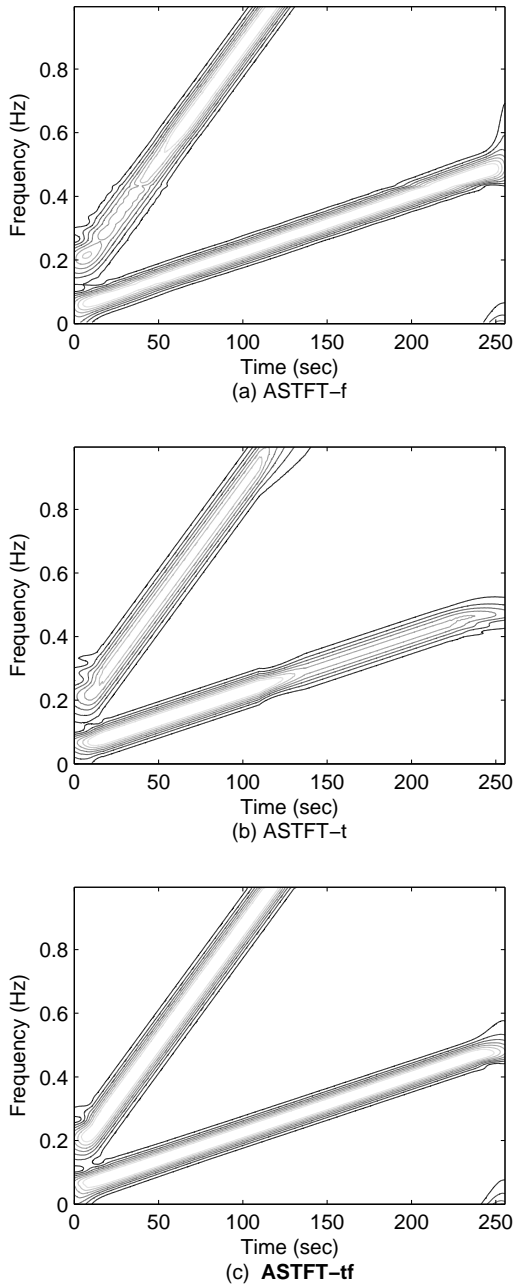}
\end{center}
\caption{The 
TFRs of a multicomponent signal: (a) ASTFT-f; (b) ASTFT-t; and (c) \textbf{ASTFT-tf}.
In this case, the ASTFT-tf has the highest energy concentration.
} \label{fig:SIM2}
\end{figure}

\mysubsection{Energy Concentration Analysis of the ASTFT-f, the ASTFT-t and the ASTFT-tf}\label{subsec:Comp}
Energy concentration of the ASTFT-f, the ASTFT-t and the
ASTFT-tf is
examined by using a multicomponent signal consisting of two linear FM components,
\begin{eqnarray*}\label{eq:SIM04}
x(t) &=& \exp \left[ {j2\pi \left( {{f_1} \cdot t + \frac{{{f_2} - {f_1}}}{{512}} \cdot {t^2}} \right)} \right] \nn\\
&+& \exp \left[ {j2\pi \left( {{f_3} \cdot t+ \frac{{{f_4} - {f_3}}}{{512}} \cdot {t^2}} \right)} \right],
\end{eqnarray*}
where ${f_1} = 0.05,{\rm{ }}{f_2} = 0.5,{\rm{ }}{f_3} = 0.15,{\rm{ }}{f_4} = 2$.
Fig.~\ref{fig:SIM2} shows 
these three TFRs of the signal with $\Delta_t=1/2$ and $\Delta_f=1/256$.
The energy concentration of the \textbf{ASTFT-tf} is higher than that of the ASTFT-f.
This is because the standard deviation of the Gaussian kernel in the ASTFT-f is time-independent.
Therefore, observing the 
ASTFT-f from $f=0.15$ to $f=0.5$ in Fig.~\ref{fig:SIM2}(a), the obtained standard deviation cannot be simultaneously the optimal for both the components.
Similarly, the \textbf{ASTFT-tf} has higher energy concentration than the ASTFT-t since the standard deviation in the ASTFT-t is frequency-independent.
Thus, observing the 
ASTFT-t from $t=0$ to $t=120$ in Fig.~\ref{fig:SIM2}(b), the obtained standard deviation can not be simultaneously the optimal for both the components.
Since these two components have different chirp rates, it is better to use time-frequency-varying standard deviation adapted to each component.
This example verifies that
the \textbf{ASTFT-tf} is superior to the ASTFT-f and the ASTFT-t 
for signals having multiple chirp rates at some time instant or frequency.

\begin{figure}
\vspace{0.5cm}\begin{center}
\includegraphics[width=6.75cm]{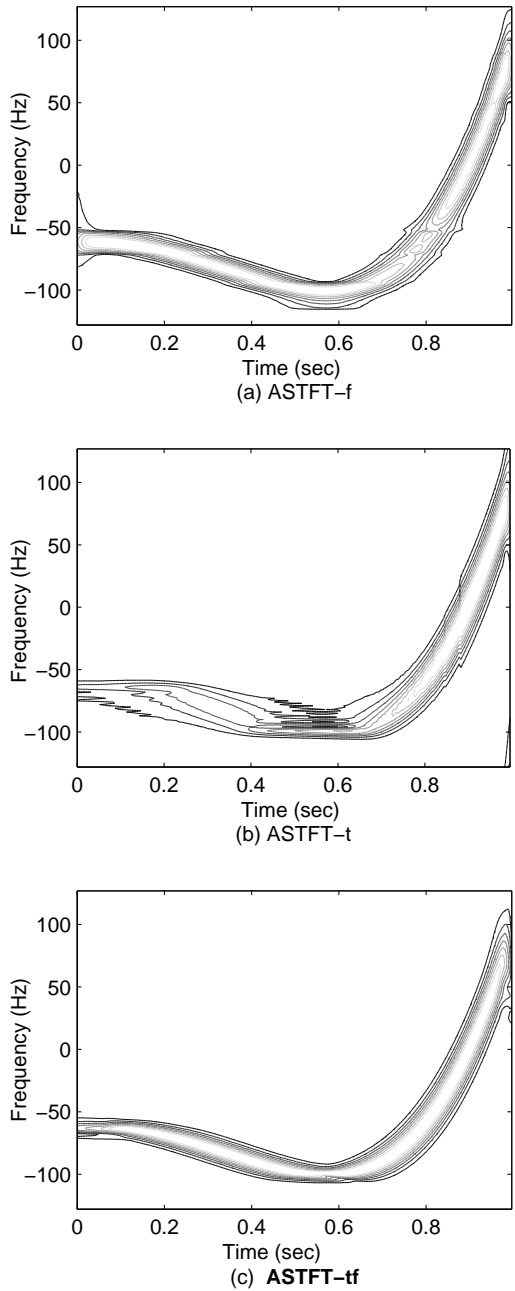}
\end{center}
\caption{The 
TFRs of a nonlinear FM signal: (a) ASTFT-f; (b) ASTFT-t; and (c) \textbf{ASTFT-tf}.
In this case, the ASTFT-tf has the highest energy concentration.
} \label{fig:SIM3}
\end{figure}

Consider another signal which comprises one nonlinear FM component,
\vsp{-0.1cm}
\begin{equation*}\label{eq:SIM06}
x(t) = \exp \left( {j2\pi (100{t^5} - 25{t^4} - 85{t^3} + 8{t^2} - 62t)} \right).
\vsp{-0.1cm}\end{equation*}
The ASTFT-f, the ASTFT-t and the
ASTFT-tf of the signal with $\Delta_t=1/256$ and $\Delta_f=1$ are depicted in Fig.~\ref{fig:SIM3}.
At any frequency between $-102$Hz and $-62$Hz, there are two different chirp rates along the time axis.
Therefore, the ASTFT-f is
no doubt inferior to the \textbf{ASTFT-tf} in this frequency band.
It has been illustrated in Fig.~\ref{fig:mask3} and Section \ref{subsec:Interp} that the time-frequency-varying standard deviation is still a better choice even though the signal has single chirp rate at any time instant or frequency.
Therefore, at any frequency larger than $-62$Hz in Figs.~\ref{fig:SIM3}(a) and \ref{fig:SIM3}(c), it can be found that the \textbf{ASTFT-tf} somewhat outperforms the ASTFT-f.
From Fig.~\ref{fig:SIM3}(b), it is shown that the ASTFT-t suffers from poor energy concentration for two main reasons: first, the standard deviation is time-varying but not frequency-varying; second, the relationship between the standard deviation and the chirp rate in (\ref{eq:IFG04}) is not adequate.
Besides, as mentioned in Section~\ref{subsec:IFG}, there's no criterion for determining the threshold $\xi$ used in this relationship.
$\xi=0.07$ is used for the signal in Fig.~\ref{fig:SIM2}, while $\xi=25$ is applied to the signal in Fig.~\ref{fig:SIM3}.
These values of $\xi$ are obtained by means of try and error such that most part of the 
ASTFT-t has high energy concentration.
In contrast, the energy 
distribution in Fig.~\ref{fig:SIM3}(c) shows that the nonparametric relationship used in the \textbf{ASTFT-tf} is capable of achieving much higher energy concentration, even though the ASTFT-t and ASTFT-tf use the 
same estimated chirp rate.

\begin{figure*}[t]
\begin{center}
\includegraphics[width=14cm]{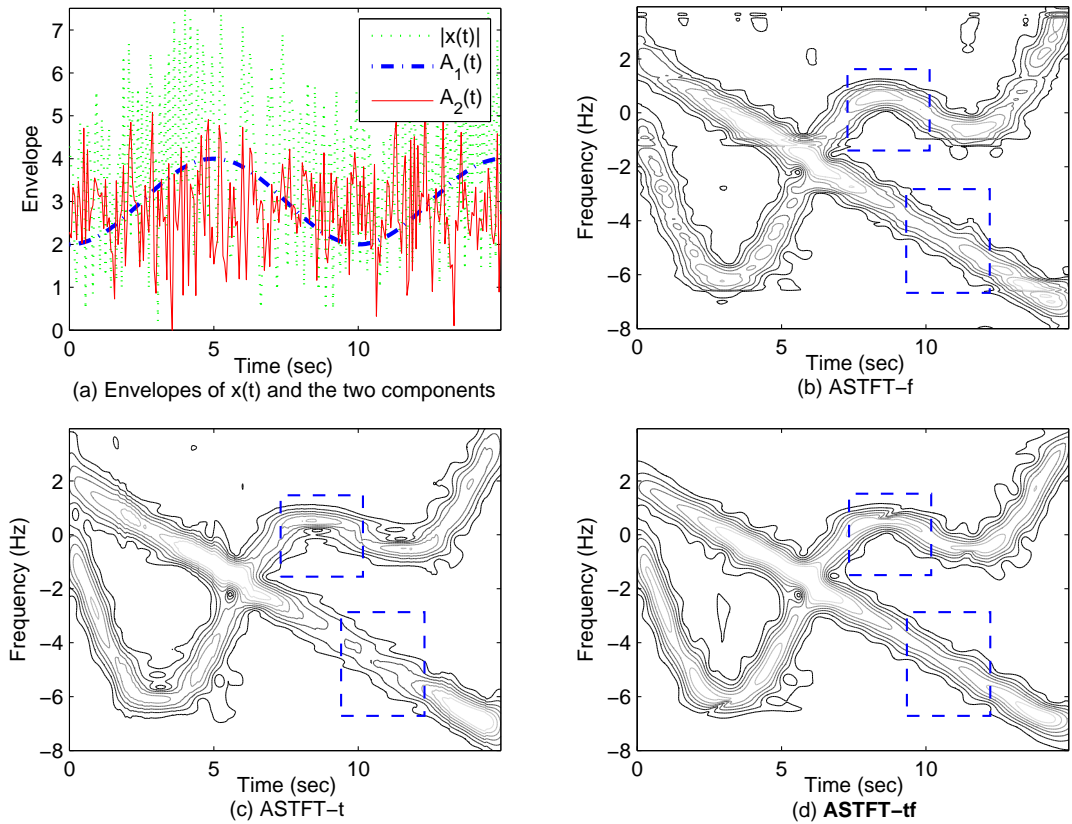}
\end{center}
\vspace{-0.3cm}\vsp{-0.5cm}\caption{
TFRs of a more general multicomponent signal: (a) envelopes of the signal $x(t)$ and its two components ($A_1(t)$ is the sinusoidal envelope of the linear FM component, while $A_2(t)$ is the positive random envelope of the nonlinear component); (b) ASTFT-f; (c) ASTFT-t; and (d) \textbf{ASTFT-tf}.
The ASTFT-tf is somewhat better than the ASTFT-f and the ASTFT-t, especially for the TF regions within the dashed rectangles.
} \label{fig:SIM_AQ}
\vspace{-0.3cm}\vsp{-0.7cm}\end{figure*}

Considering the more general signal model $x(t) = \sum\limits_k {{A_k}} (t)\exp \left( {j{\varphi _k}(t)} \right)$ where ${A_k}(t) \ge 0$, another simulation result is given in Fig.~\ref{fig:SIM_AQ}.
In this simulation, the signal under analysis consists of a linear FM component with sinusoidal envelope and a nonlinear FM component with positive random envelope,
\begin{eqnarray*}\label{eq:SIM07}
x(t) &=& {A_1}(t)\exp \left( {j2\pi \left( { - {\rm{0}}{\rm{.3125}}{t^2} + 2t} \right)} \right)\nn\\
&+& {A_2}(t)\exp \left( {j2\pi \left( {13\cos 0.1\pi t + 5\cos 0.2\pi t} \right)} \right),
\end{eqnarray*}
where ${A_1}(t) =  - \cos 0.2\pi t + 3$ and $A_2(t)$ is the absolute value of a Gaussian random signal with unit variance.
The envelope of  $x(t)$, $A_1(t)$  and $A_2(t)$  are shown in Fig.~\ref{fig:SIM_AQ}(a). The ASTFT-f, the ASTFT-t and the ASTFT-tf of the signal with ${\Delta _t} = 1/16$  and  ${\Delta _f} = 1/16$ are depicted in Figs.~\ref{fig:SIM_AQ}(b), \ref{fig:SIM_AQ}(c) and \ref{fig:SIM_AQ}(d), respectively. The ASTFT-tf is somewhat better than the ASTFT-f and the ASTFT-t, especially for the TF regions within the dashed rectangles shown in Fig.~\ref{fig:SIM_AQ}.

\mysubsection{Energy Concentration Analysis of the FFT-based ASTFT-tf}\label{subsec:CompFFT}
The FFT can be used in the ASTFT-f \cite{Pei} and the ASTFT-t \cite{Zhong}. 
The precondition of using FFT in the ASTFT-tf is that $\sigma [m,n]$ is irrelevant to $m$ or $n$.
Interested readers can refer to APPENDIX~\ref{App:B} for details of the FFT-based implementation of the ASTFT-tf.
Now the problem is how to determine the $\sigma [m]$ (or $\sigma [n]$) when encountering multiple chirp rates.
A simple and straightforward 
approach is to average the chirp rates along the frequency axis (i.e. $\AW{f'}_{inst}[m]$) or along the time axis (i.e. $\AW{f'}_{inst}[n]$).
The former leads to $\sigma [m,n]=\sigma [m]$ while the latter yields $\sigma [m,n]=\sigma [n]$.
Chirp rate interpolation discussed at the end of Section~\ref{subsec:Interp} is unnecessary.
The cost of using averaged chirp rate is the performance loss because the standard deviation is no longer 
time-frequency-varying.
The choice between using $\sigma [m,n]=\sigma [m]$ and using $\sigma [m,n]=\sigma [n]$ is dependent 
on the chirp rates of the signal.
For each time point, if the absolute values of all the chirp rates are close to each other, 
the chirp rates are averaged along the frequency axis.
For each frequency point, if the absolute values of all the chirp rates are close to each other, 
chirp rates are averaged along the time axis.

\begin{figure*}[t]
\begin{center}
\includegraphics[width=14cm]{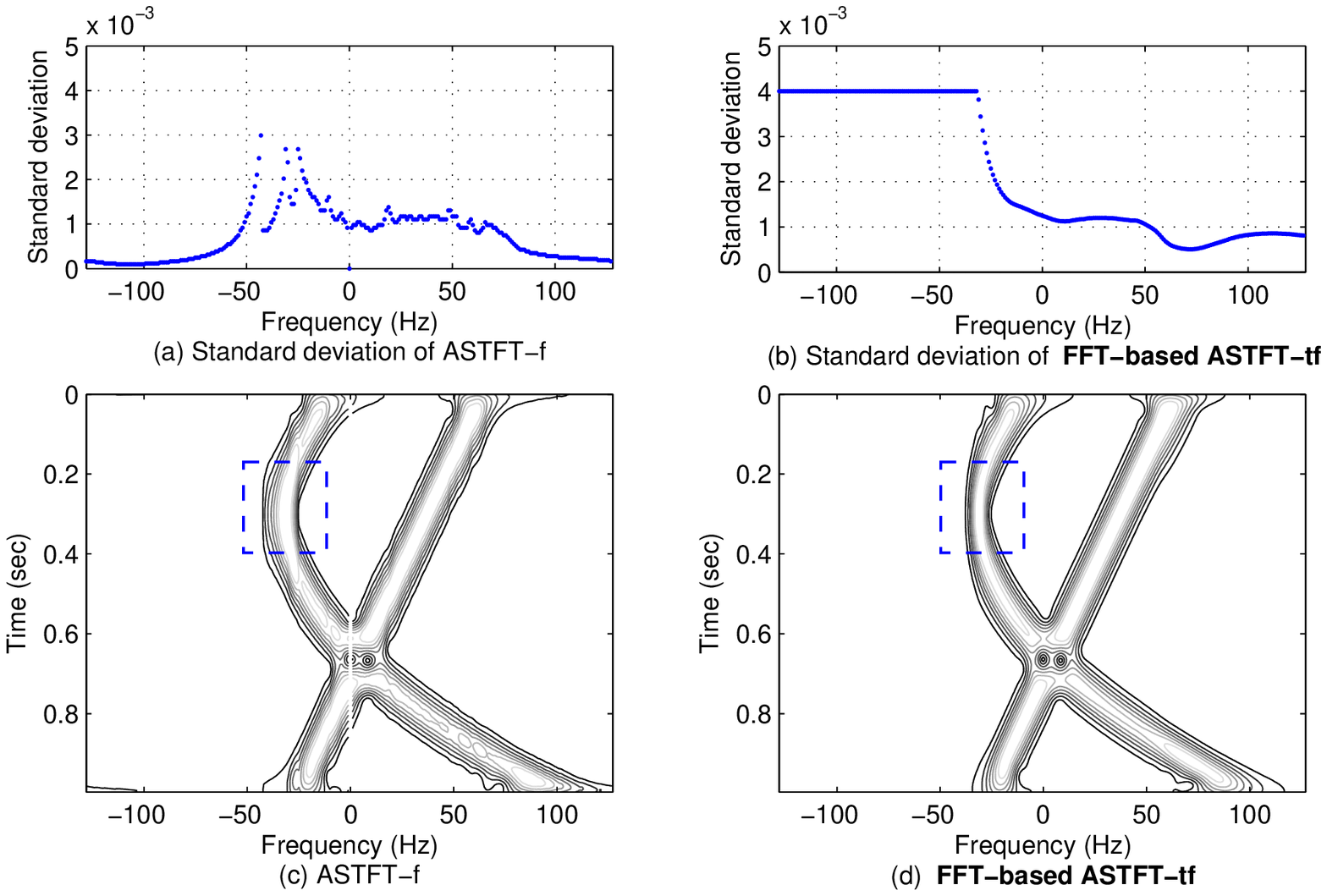}
\end{center}
\vspace{-0.3cm}\vsp{-0.5cm}\caption{
TFRs of a multicomponent signal and the frequency-varying standard deviations used in these TFRs (note that the horizontal axis represents the frequency): (a) standard deviation used in the ASTFT-f; (b) standard 
deviation used in the \textbf{FFT-based ASTFT-tf}; (c) ASTFT-f; and (d) \textbf{FFT-based ASTFT-tf} (based on averaging the chirp rates along the time axis).
The FFT-based ASTFT-tf is somewhat more concentrated than the ASTFT-f, especially near $\{t=0.3,f=-32\}$ (within the dashed rectangle).
} \label{fig:SIM4f}
\vspace{-0.3cm}\vsp{-0.7cm}\end{figure*}
\begin{figure*}[t]
\begin{center}
\includegraphics[width=14cm]{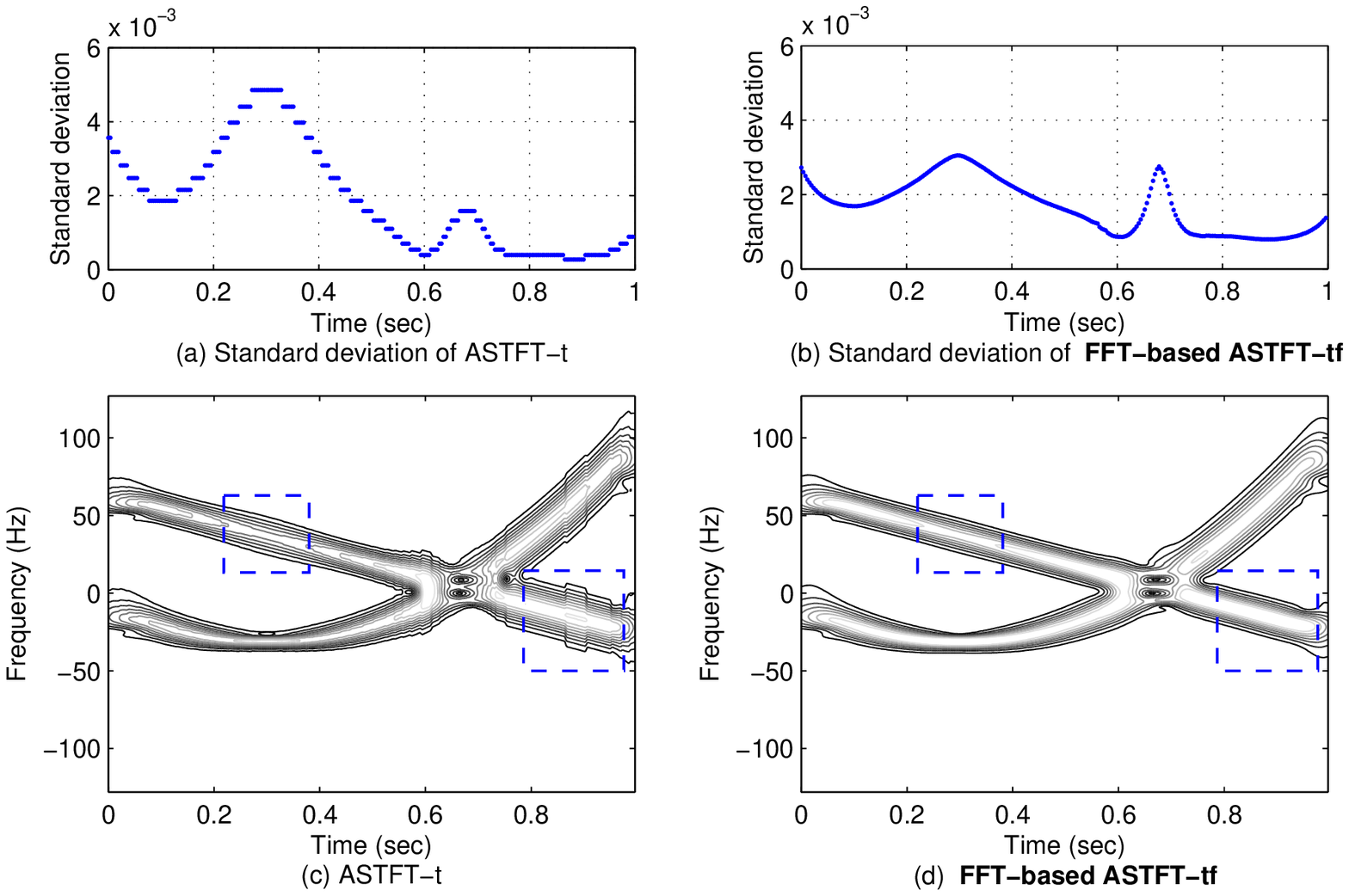}
\end{center}
\vspace{-0.4cm}\vsp{-0.5cm}\caption{
TFRs of a multicomponent signal and the time-varying standard deviations used in these TFRs (note that the horizontal axis represents the time): (a) standard deviation used in the ASTFT-t; (b) standard 
deviation used in the \textbf{FFT-based ASTFT-tf}; (c) ASTFT-t; and (d) \textbf{FFT-based ASTFT-tf} (based on averaging the chirp rates along the frequency axis).
The FFT-based ASTFT-tf has higher energy concentration than the ASTFT-t, especially near 
$\{t=0.3,f=-32\}$ and $\{t=0.9,f=-17\}$ (within the dashed rectangles).
} \label{fig:SIM4t}
\vsp{-0.6cm}\end{figure*}

The comparison between the ASTFT-t, the ASTFT-f and the
FFT-based ASTFT-tf is presented using a synthetic multicomponent signal
\begin{eqnarray*}\label{eq:SIM18}
x(t) &=& \exp \left( {j2\pi \left( {90{{(t - 0.3)}^3} - 32t} \right)} \right)\nn\\
 &+& \exp \left( {j2\pi \left( { - 45{t^2} + 64t} \right)} \right).
\end{eqnarray*}
Figs.~\ref{fig:SIM4f}(c) and \ref{fig:SIM4f}(d) 
show the ASTFT-f and the
FFT-based ASTFT-tf 
(using standard deviation based on averaging the chirp rates along the time axis).
Figs.~\ref{fig:SIM4f}(a) and \ref{fig:SIM4f}(b) 
show the corresponding
frequency-varying standard deviations adopted in Figs.~\ref{fig:SIM4f}(c) and \ref{fig:SIM4f}(d), respectively.
Note that the horizontal axis of the figures represents the frequency.
The synthetic signal occupies frequency band from $-32$Hz to $100$Hz.
The \textbf{FFT-based ASTFT-tf} is somewhat superior to the ASTFT-f, especially near $\{t=0.3,f=-32\}$ as shown in Figs.~\ref{fig:SIM4f}(c) and \ref{fig:SIM4f}(d).
The reason is that the CM2 optimization (referring to (\ref{eq:CM10})) in the ASTFT-f is to maximize the ``total'' energy concentration along the time axis.
Therefore,
even though the ASTFT-f has the highest CM2 (the standard deviation in the ASTFT-f is the optimal in the sense of CM2),
an undesirable phenomenon may be induced: some components may have much higher energy concentration while some others may have much lower energy concentration.
In contrast, the FFT-based ASTFT-tf can maximize the energy concentration of all the components more fairly because the the standard deviation is dependent on the ``averaged'' chirp rate.
Fig.~\ref{fig:SIM4t} depicts 
the ASTFT-t and the
FFT-based ASTFT-tf 
(using standard deviation based on averaging the chirp rates along the frequency axis) and the time-varying standard deviations used in these two methods.
Note that the horizontal axis of the figures  represents the time.
The standard deviations are similar to each other, but the ASTFT-t is inferior to the
\textbf{FFT-based ASTFT-tf}, especially near $\{t=0.3,f=-32\}$ and $\{t=0.9,f=-17\}$.
This is because the relationship for calculating the standard deviation in the ASTFT-t is not adequate.
This relationship is derived for the purpose of quasi-stationarity rather than maximizing the energy concentration.

Besides energy concentration, the other significant advantages of the proposed methods should also be kept in mind: the robustness of the chirp rate estimator, adaptivity and complexity. Therefore, a detailed comparison among these ASTFTs is given in TABLE~\ref{tab:table1}.
Compared with the \textbf{FFT-based ASTFT-tf}, the ASTFT-f is also completely adaptive to the signal, but has much higher complexity due to its optimization process.
The ASTFT-t in all the simulations uses the estimated chirp rate obtained from the \textbf{ASTFT-tf}, but is less adaptive to the signal, because the non-adaptive threshold $\xi$ in (\ref{eq:IFG04}) and (\ref{eq:IFG08}) is signal-dependent.
Also, it has somewhat higher complexity than the FFT-based ASTFT-tf, because the calculation of the standard deviation in the FFT-based ASTFT-tf is simpler than that in the ASTFT-t.
Besides the non-adaptive threshold  $\xi$, the \textbf{original} ASTFT-t (the original method proposed in \cite{Zhong}) uses the non-adaptive WT for IF estimation, and thus is much less adaptive to the signal. Furthermore, it is less robust to IF estimation error, because the difference operator is employed for chirp rate estimation. Because of the difference operator and the non-adaptive WT, it has lower complexity for tradeoff of much lower energy concentration.
\begin{table*}
\begin{center}
\setstretch{1.5}
\caption{Comparisons of chirp rate estimator, adaptivity, complexity and energy concentration.}\label{tab:table1}
\begin{tabular}{|c|c|c|c|c|}
\hline
\multirow{3}{*}{TFR} &Chirp rate & \multirow{2}{*}{Adaptivity} & \multirow{2}{*}{Complexity}  & Energy\\
 & estimation &   &    & concentration\\
\cline{2-5}
 & \multicolumn{4}{|c|}{(w.r.t. \textbf{FFT-based ASTFT-tf})}  \\
\hline\hline
ASTFT-f \cite{Pei}  & Not used & Same & Much higher & A little lower\\
\hline
\footnotemark[3]Original ASTFT-t \cite{Zhong}  &Less robust & Much less adaptive & Lower  & Much lower\\
\hline
\footnotemark[3]ASTFT-t & Same & Less adaptive & A little higher  & Lower\\
\hline
\textbf{ASTFT-tf}  & Same & Same & Higher & Higher\\
\hline
\end{tabular}
\end{center}
\vsp{-0.8cm}
\vspace{-0.5cm}
\end{table*}
%

\vsp{-0.2cm}
\mysubsection{Comparison of ASTFT-tf with other Adaptive TF Representations in Noisy Environments}\label{subsec:Noise}
In this subsection, the performance of a variety of adaptive TFRs is examined in noisy environments.
To  design an adaptive TFR, the approaches considered here include the CMs and the reassignment methods.
TFRs considered here are the STFT (adaptive S-transform is a special case of ASTFT) and some popular bilinear TFRs including the SPWVD and the SM.
These bilinear TFRs do not have cross-term problem, and the CM approach and the reassignment methods can be easily applied to them.
\footnotetext[3]{The \textbf{original} ASTFT-t is the original method proposed in \cite{Zhong}, using the WT for IF estimation and the difference operator for chirp rate evaluation. The ASTFT-t uses the same IF and chirp rate estimators as in the ASTFT-tf (i.e. using the CM5-based ASTFT for IF estimation and the PCA to calculate the chirp rate).}
In the following, the performance of seven adaptive TFRs are compared:
ASTFT-tf (our method), ASTFT-t \cite{Zhong} (using chirp rate obtained from the ASTFT-tf), CM-based STFT (i.e. the ASTFT-f) \cite{Pei}, CM-based SM (CM-SM) \cite{Stankovic2}, CM-based SPWVD (CM-SPWVD) \cite{Behzad}, reassigned SM (RSM) \cite{Djurovi2}, and reassigned SPWVD (RSPWVD) \cite{Auger}.
Matlab code of the RSPWVD is available in \cite{Auger2}.

\begin{figure*}
\begin{center}
\includegraphics[width=14cm]{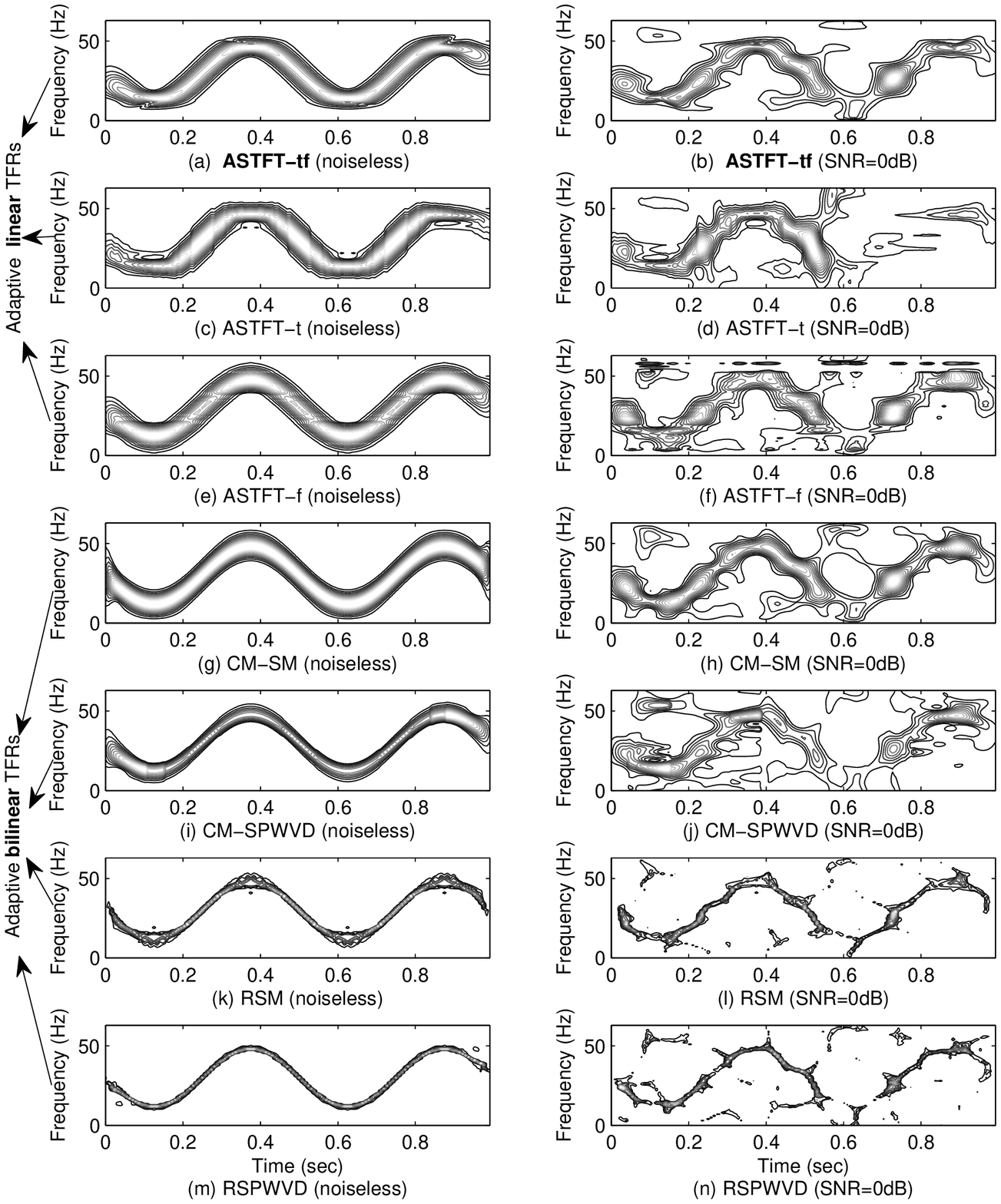}
\end{center}
\vsp{-0.9cm}\caption{
(a), (c), (e), (g), (i), (k) and (m) are TFRs in noiseless environments, while (b), (d), (f), (h), (j), (l) and (n) are TFRs in noisy environments with SNR=0dB. Seven adaptive TFRs are concerned: \textbf{ASTFT-tf} in (a) and (b); ASTFT-t in (c) and (d); ASTFT-f in (e) and (f); CM-SM in (g) and (h); CM-SPWVD in (i) and (j); RSM in (k) and (l); RSPWVD in (m) and (n).
In noiseless condition, \textbf{ASTFT-tf} is more concentrated than ASTFT-t and ASTFT-f but less concentrated than most of the adaptive bilinear TFRs.
In noisy condition, \textbf{ASTFT-tf} still outperforms ASTFT-t; CM-based and reassigned TFRs enhance the energy concentration for all the ``noise-only'' and `signal+noise'' components.
} \label{fig:Noise}
\vsp{-0.5cm}\end{figure*}

Consider the signal $x(t) = \exp \left( {j60\pi t+3\pi\cos(4\pi t)} \right)$.
Fig.~\ref{fig:Noise} depicts the seven adaptive TFRs with $\Delta_t=1/128$ and $\Delta_f=1$ in
both noiseless and noisy (SNR=0dB) environments.
In noiseless environments,
the RSM and the RSPWVD have the highest concentration level.
Among the CM-based TFRs, the CM-SPWVD outperforms the ASTFT-f and the CM-SM, because it is bilinear and its CM optimization algorithm is more complicated than that in the CM-SM.
Generally, the ASTFTs are inferior to the adaptive bilinear TFRs in noiseless environments.
In noisy environments, the RSM and the RSPWVD 
enhance energy concentration for all the components,
and it becomes more difficult to distinguish 
the ``noise-only'' and the ``signal+noise'' components, especially at low SNR.
The CM-based methods are also sensitive to noise because they enhance the total energy concentration 
of all the components.
The worst situation is that the optimal standard deviation yields high concentration 
for the ``noise-only'' components but low concentration for the ``signal+noise'' components.
In contrast, chirp-rate-based methods (i.e. the \textbf{ASTFT-tf} and the ASTFT-t) are not affected by noise directly.
Noise affects the accuracy of the estimated chirp rate, and then the estimation error affects the performance of the chirp-rate-based methods.
Fortunately, our chirp rate estimator (refer to Section~\ref{subsec:CRE}) is somewhat robust to the estimation error.

\begin{figure}[t]
\begin{center}
\includegraphics[width=8.85cm]{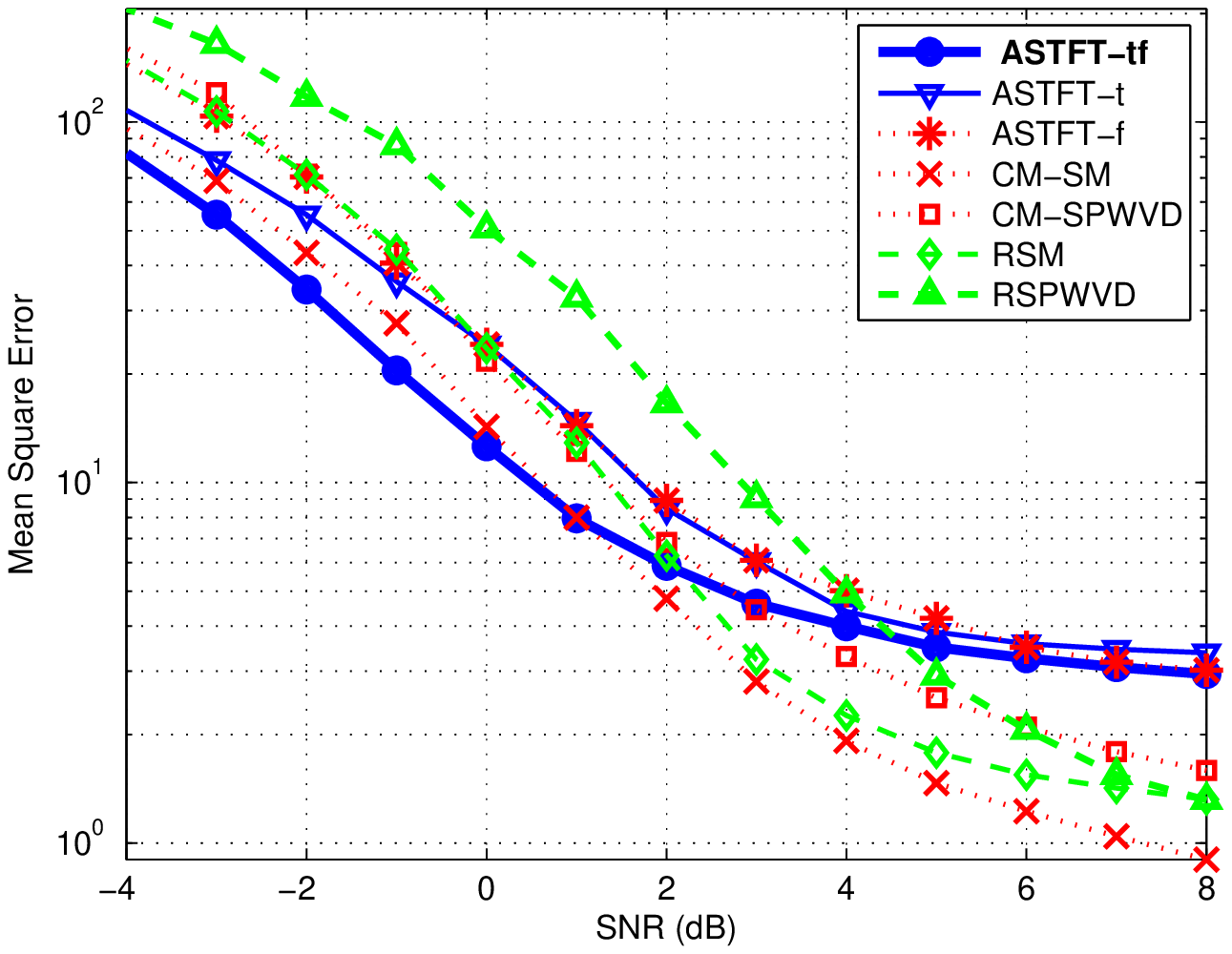}
\end{center}
\vsp{-0.5cm}\caption{
MSE of the IF estimation based on \textbf{ASTFT-tf}, ASTFT-t, ASTFT-f, CM-SM, CM-SPWVD, RSM, and RSPWVD.
The \textbf{ASTFT-tf} (thick solid line) is superior to all the other adaptive TFRs at low SNR but inferior to the adaptive bilinear TFRs at high SNR.
} \label{fig:MSE}
\vsp{-0.6cm}\end{figure}

Fig.~\ref{fig:MSE} depicts SNR versus the mean squared error (MSE) of IF estimation
based on the seven adaptive TFRs.
The MSE is defined as
\vsp{-0.2cm}\begin{equation*}\label{eq:Noise01}
E\left\{ \frac{1}{N}\sum\limits_{n = 0}^{N - 1} {|{{\AF f'}_{inst}}[n] - {{f'}_{inst}}[n]{|^2}} \right\},
\end{equation*}
where  ${{f'}_{inst}}[n]$ and ${{\AF f'}_{inst}}[n]$ are respectively the exact and estimated chirp rates.
The ASTFT-tf is superior to all other adaptive TFRs at low SNR but inferior to the adaptive bilinear TFRs at high SNR. However, in some applications such as signal analysis and synthesis, the ASTFT-tf may be more useful in both low SNR and high SNR environments because it is a linear transform.

\mysection{Conclusion}\label{sec:Con}
In this paper, the chirp-rate-based ASTFT presented in \cite{Zhong} has been substantially modified.
First, because the 
wavelet transform (WT) used for IF estimation is not signal-dependent, 
a low-complexity ASTFT based on a novel CM
has been designed.
Second, instead of using the difference operator to calculate the chirp rate,
a more robust chirp rate estimator has been proposed.
This robust mechanism eliminates some IF estimation error and uses the PCA to calculate the chirp rate for the robustness 
to the remaining IF estimation error.
Third, based on the approximate relationship between the optimal time-varying window width and the chirp rate derived by Cohen \cite{Cohen}, 
a Gaussian kernel with time-frequency-varying window width have been introduced, which is more suitable for nonlinear FM signals and multicomponent signals.
Based on these modifications, a novel chirp-rate-based ASTFT (called \textbf{ASTFT-tf}) and the \textbf{FFT-based ASTFT-tf} have been proposed.
The
ASTFT-tf inherits the benefit of the chirp-rate-based ASTFT that the complexity is much lower than that in the CM-based ASTFT.
Simulation results show that 
the
ASTFT-tf has higher energy concentration than the CM-based and the chirp-rate-based ASTFTs.
Also, for IF estimation, it has been shown that 
the
ASTFT-tf is superior to many other adaptive TFRs at low SNR but inferior to the adaptive bilinear TFRs at high SNR. However, in some applications such as signal analysis and synthesis, 
the
ASTFT-tf may be more useful for both low SNR and high SNR conditions because it is a linear transform.

\appendices
\section{Envelope of the ASTFT-tf of a linear FM signal}\label{App:A}
Consider the signal under analysis is a linear FM signal of the form $x(t) = \exp \left( {j2\pi (a{t^2}/2 + bt)} \right)$.
If $\sigma(t,f)=\sigma$, the ASTFT-tf of the signal is given by
\begin{eqnarray}\label{eq:A0}
&&ASTFT_{tf}(t,f)\nn\\
&=& \int\limits_{ - \infty }^\infty  {{e^{j2\pi \left( {\frac{a}{2}{\tau ^2} + b\tau } \right)}}\frac{1}{{\sqrt {2\pi } \sigma }}{e^{ - \frac{{{{(t - \tau )}^2}}}{{2{\sigma ^2}}}}}{e^{ - j2\pi f\tau }}d\tau } \nn\\
&=& \frac{1}{{\sqrt {2\pi } \sigma }}{e^{ - \frac{{{t^2}}}{{2{\sigma ^2}}}}}\int\limits_{ - \infty }^\infty  {{e^{j\pi a{\tau ^2} - \frac{1}{{2{\sigma ^2}}}{\tau ^2} + j2\pi b\tau  + \frac{t}{{{\sigma ^2}}}\tau  - j2\pi f\tau }}d\tau }.\qquad
\end{eqnarray}
According to \cite{Ryzhik}, if ${\mathop{\rm Re}\nolimits} \{ {\mu ^2}\}  > 0$,
\begin{equation}\label{eq:A2}
\int_{ - \infty }^\infty  {e^ { - {\mu ^2}{x^2} \pm \nu x}} dx = \frac{{\sqrt \pi  }}{\mu }e^{{\frac{{{\nu ^2}}}{{4{\mu ^2}}}} }.
\end{equation}
Assume $\mu  = \sqrt {\frac{1}{{2{\sigma ^2}}} - j\pi a} $ and $\nu  = \frac{t}{{{\sigma ^2}}} - j2\pi (f - b)$, and then (\ref{eq:A0}) can be simplified as
\begin{equation}\label{eq:A4}
ASTFT_{tf}(t,f) = \frac{1}{{\sqrt {2\pi } \sigma }}{e^{ - \frac{{{t^2}}}{{2{\sigma ^2}}}}}\frac{{\sqrt \pi  }}{\mu }{e^{\left( {\frac{{{\nu ^2}}}{{4{\mu ^2}}}} \right)}}{\rm{ }} = {c_0}{e^{{c_1}}},
\end{equation}
where
\begin{eqnarray}\label{eq:A6}
{c_0} &=&\frac{1}{{\sqrt {2\pi } \sigma }}\cdot\frac{{\sqrt \pi  }}{\mu } = \frac{1}{{\sqrt {1 - j2\pi {\sigma ^2}a} }},\nn\\
{c_1} &=&  - \frac{{{t^2}}}{{2{\sigma ^2}}} + \frac{{{{\left( {t - j2\pi {\sigma ^2}(f - b)} \right)}^2}}}{{2{\sigma ^2}\left( {1 - j2\pi {\sigma ^2}a} \right)}}.
\end{eqnarray}
To determine the envelope of the ASTFT-tf,the real part of $c_1$ is evaluated,
\begin{eqnarray}\label{eq:A8}
&&{\mathop{\rm Re}\nolimits} \{ {c_1}\}\nn\\
&&\quad=  - \frac{1}{{2{\sigma ^2}}}\left[ {{t^2} - \frac{{{t^2} - 4{\pi ^2}{\sigma ^4}{{(f - b)}^2} + 8{\pi ^2}{\sigma ^4}a(f - b)t}}{{1 + 4{\pi ^2}{\sigma ^4}{a^2}}}} \right]\nn\\
&&\quad=  - \frac{{2{\pi ^2}{\sigma ^2}}}{{1 + 4{\pi ^2}{\sigma ^4}{a^2}}}{\left( {f - b - at} \right)^2}.
\end{eqnarray}
Therefore, the envelope is
\begin{eqnarray}\label{eq:A10}
&&\left| ASTFT_{tf}(t,f) \right| = \left| {{c_0}} \right|{e^{{\mathop{\rm Re}\nolimits} \{ {c_1}\} }}\nn\\
&&\qquad\qquad=\left( {1 + 4{\pi ^2}{\sigma ^4}{a^2}} \right)^{ - \frac{1}{4}}}{e^{ - \frac{{2{\pi ^2}{\sigma ^2}}}{{1 + 4{\pi ^2}{\sigma ^4}{a^2}}}{{\left( {f - b - at} \right)}^2}}.\quad
\end{eqnarray}

\section{FFT-based Implementation of the ASTFT-tf}\label{App:B}
Recall the aforementioned discrete ASTFT-tf in (\ref{eq:ASTFT04}).
Assume $\sigma _{\max }$ is the upper bound of all the $\sigma [m,n]$'s, and then
\begin{equation}\label{eq:SIM08}
\frac{1}{{\sqrt {2\pi } \sigma [m,n]}}{e^{ - \frac{{{m^2}\Delta _t^2}}{{2{{\left( {\sigma [m,n]} \right)}^{\rm{2}}}}}}} \le \frac{1}{{\sqrt {2\pi } {\sigma _{\max }}}}{e^{ - \frac{{{m^2}\Delta _t^2}}{{2{{\left( {{\sigma _{\max }}} \right)}^{\rm{2}}}}}}}. 
\end{equation}
If the right side of the above inequality tends to $0$ as $|m|>Q_1$, the sum of infinitely many terms in (\ref{eq:ASTFT04}) can be truncated to $2Q_1+1$ terms.
That is,
\begin{eqnarray}\label{eq:SIM10}
&& \sum\limits_{l = m - Q_1}^{m + Q_1} {x[l]\frac{1}{{\sqrt {2\pi } \sigma [m,n]}}{e^{ - \frac{{{{(m - l)}^2}\Delta _t^2}}{{2{{\left( {\sigma [m,n]} \right)}^{\rm{2}}}}}}}{e^{ - j2\pi nl{\Delta _t}{\Delta _f}}}{\Delta _t}} \nn\\
 &=& {\Delta _t}\sum\limits_{l = 0}^{2Q_1} {\frac{{x[m - {Q_1} + l]}}{{\sqrt {2\pi } \sigma [m,n]}}{e^{ - \frac{{{{({Q_1} - l)}^2}\Delta _t^2}}{{2{{\left( {\sigma [m,n]} \right)}^{\rm{2}}}}}}}{e^{ - j\frac{{2\pi n(m - {Q_1} + l)}}{N}}}}  \nn\\
 &=& {\Delta _t}{e^{ - j\frac{{2\pi n(m - Q_1)}}{N}}}\sum\limits_{l = 0}^{N - 1} {{x_1}[m,n,l]{e^{ - j2\pi \frac{{nl}}{N}}}},
\end{eqnarray}
where $N = 1/({\Delta _t}{\Delta _f}) \geq 2Q_1 + 1$ and
\begin{equation*}\label{eq:SIM11}
{x_1}[m,n,l] = \left\{ {\begin{array}{*{20}{c}}
{\frac{{x[m - {Q_1} + l]}}{{\sqrt {2\pi } \sigma [m,n]}}{e^{ - \frac{{{{({Q_1} - l)}^2}\Delta _t^2}}{{2{{\left( {\sigma [m,n]} \right)}^{\rm{2}}}}}}},}&{\;0 \le l \le 2{Q_1}}\\
{0,}&{\;2{Q_1} < l < N}
\end{array}} \right..
\end{equation*}
It is apparent that the FFT can be applied to (\ref{eq:SIM10}) if $x_1[m,n,l]$ is irrelevant to $n$; that is, ${\sigma [m,n]}$ does not change with $n$.

Using the notion of the FFT implementation in the S-transform, the discrete ASTFT-tf can also be expressed by
\begin{eqnarray}\label{eq:SIM12}
&&ASFT_{tf}[m,n]\nn\\
&&\;= \sum\limits_{k =  - \infty }^\infty  {X[k + n]\; {e^{ - 2{\pi ^2}{{\left( {\sigma [m,n]} \right)}^2}{k^2}\Delta _f^2}}{e^{j2\pi mk{\Delta _t}{\Delta _f}}}{\Delta _f}},\qquad
\end{eqnarray}
where $X[k]$ is the DFT of the $x[l]$.
It is obvious that (\ref{eq:SIM12}) is equivalent to the discrete S-transform as $\sigma [m,n] = 1/(n{\Delta _f})$.
Assume $\sigma _{\min }$ is the lower bound of all the $\sigma [m,n]$'s, and then we have
\begin{equation*}\label{eq:SIM14}
{e^{ - 2{\pi ^2}{{\left( {\sigma [m,n]} \right)}^2}{k^2}\Delta _f^2}} \le {e^{ - 2{\pi ^2}{{\left( {{\sigma _{\min }}} \right)}^2}{k^2}\Delta _f^2}}. 
\end{equation*}
If the right side of the above inequality tends to $0$ as $|k|>Q_2$, the sum of infinitely many terms in (\ref{eq:SIM12}) can be truncated to $2Q_2+1$ terms.
That is,
\begin{eqnarray}\label{eq:SIM16}
&&\sum\limits_{k =  - {Q_2}}^{{Q_2}} {X[k + n]\;{\mkern 1mu} {e^{ - 2{\pi ^2}{{\left( {\sigma [m,n]} \right)}^2}{k^2}\Delta _f^2}}{e^{j2\pi mk{\Delta _t}{\Delta _f}}}{\Delta _f}} \nn\\
 &=& {\Delta _f}\sum\limits_{k = 0}^{2Q_2} {X[k - {Q_2} + n]\;{\mkern 1mu} {e^{ - 2{\pi ^2}{{\left( {\sigma [m,n]} \right)}^2}{{(k - Q)}^2}\Delta _f^2}}{e^{j\frac{{2\pi m(k - Q)}}{N}}}} \nn\\
 &=& {\Delta _f}{e^{ - j\frac{{2\pi mQ}}{N}}}\sum\limits_{k = 0}^{N - 1} {{X_1}[m,n,k]\;{e^{j\frac{{2\pi mk}}{N}}}} ,\qquad\qquad
\end{eqnarray}
where $N = 1/({\Delta _t}{\Delta _f}) \geq 2Q_2 + 1$ and
\begin{equation*}\label{eq:SIM17}
{X_1}[m,n,l] = \left\{ {\begin{array}{*{20}{c}}
\begin{array}{l}
X[k - {Q_2} + n]\;{\kern 1pt} \\
 \; \times {e^{ - 2{\pi ^2}{{\left( {\sigma [m,n]} \right)}^2}{{(k - Q)}^2}\Delta _f^2}},
\end{array}&{0 \le k \le 2{Q_2}}\\
{0,}&{2{Q_2} < k < N}
\end{array}} \right..\quad
\end{equation*}
It is apparent that the FFT can be used in (\ref{eq:SIM16}) if $X_1[m,n,l]$ is irrelevant to $m$; that is, ${\sigma [m,n]}$ does not change with $m$.


\IEEEtriggeratref{38}

\bibliographystyle{IEEEtran}


\begin{IEEEbiography}[{\includegraphics[width=1in,height=1.25in,clip,keepaspectratio]{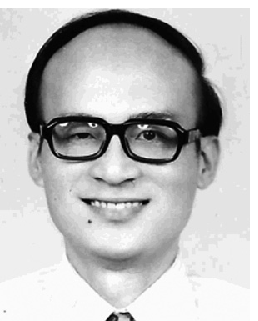}}]
{Soo-Chang Pei} (SM'89-F'00) was born in Soo-Auo, Taiwan, in 1949. He received the B.S.E.E. degree from National Taiwan University, Taipei, Taiwan, in 1970, and the M.S.E.E. and Ph.D. degrees from the University of California Santa Barbara, Santa Barbara, in 1972 and 1975, respectively. From 1970 to 1971, he was an Engineering Officer with the Chinese Navy Shipyard. From 1971 to 1975, he was a Research Assistant with the University of California Santa Barbara. He was a Professor and the Chairman of the Department of Electrical Engineering with the Tatung Institute of Technology, Taipei, from 1981 to 1983 and with National Taiwan University from 1995 to 1998. From 2003 to 2009, he was the Dean of the College of Electrical Engineering and Computer Science with National Taiwan University. He is currently a Professor with the Department of Electrical Engineering, National Taiwan University. His research interests include digital signal processing, image processing, optical information processing, and laser holography.

Dr. Pei was a recipient of the National Sun Yet-Sen Academic Achievement Award in Engineering in 1984, the Distinguished Research Award from the National Science Council from 1990 to 1998, the Outstanding Electrical Engineering Professor Award from the Chinese Institute of Electrical Engineering in 1998, the Academic Achievement Award in Engineering from the Ministry of Education in 1998, the Pan Wen-Yuan Distinguished Research Award in 2002, and the National Chair Professor Award from the Ministry of Education in 2002. He was the President of the Chinese Image Processing and Pattern Recognition Society in Taiwan from 1996 to 1998 and is a member of Eta Kappa Nu and the Optical Society of America. He became an IEEE Fellow in 2000 for his contributions to the development of digital eigenfilter design, color image coding and signal compression and to electrical engineering education in Taiwan
\end{IEEEbiography}

\begin{IEEEbiography}[{\includegraphics[width=1in,height=1.25in,clip,keepaspectratio]{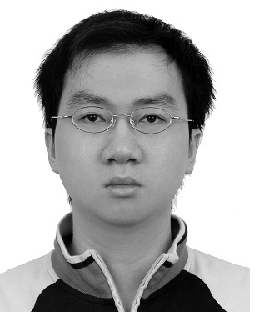}}]
{Shih-Gu Huang} received the B.S. degree in electrical engineering and the M.S. degree in communications engineering from National Tsing Hua University, Hsinchu, Taiwan, in 2007 and 2009, respectively. He is currently working toward the Ph.D. degree in the Graduate Institute of Communication Engineering, National Taiwan University, Taipei, Taiwan. His research interests include digital signal processing, time-frequency analysis, and wavelet transforms.
\end{IEEEbiography}

\vfill

\end{document}